\documentstyle[12pt]{article}

\newlength{\dinwidth}
\newlength{\dinmargin}
\def\msb{\overline{\rm MS}}

\def\GeV{{\rm GeV}}

\def\lapproxeq{\lower .7ex\hbox{$\;\stackrel{\textstyle
<}{\sim}\;$}}
\def\gapproxeq{\lower .7ex\hbox{$\;\stackrel{\textstyle
>}{\sim}\;$}}

\def\be{\begin{equation}}
\def\ee{\end{equation}}
\def\bea{\begin{eqnarray}}
\def\eea{\end{eqnarray}}
\makeatletter
\def\fmslash{\@ifnextchar[{\fmsl@sh}{\fmsl@sh[0mu]}}
\def\fmsl@sh[#1]#2{%
\mathchoice
{\@fmsl@sh\displaystyle{#1}{#2}}%
{\@fmsl@sh\textstyle{#1}{#2}}%
{\@fmsl@sh\scriptstyle{#1}{#2}}%
{\@fmsl@sh\scriptscriptstyle{#1}{#2}}}
\def\@fmsl@sh#1#2#3{\m@th\ooalign{$\hfil#1\mkern#2/\hfil$\crcr$#1
#3$}}
\makeatother

\begin{document}
\titlepage
\begin{flushright}
DTP/96/102  \\
RAL--TR--96--103 \\
December 1996 \\
(revised January 1997)
\end{flushright}

\begin{center}
\vspace*{2cm}
{\Large \bf Consistent Treatment of Charm Evolution \\[2mm]
in Deep Inelastic Scattering} \\
\vspace*{1cm}
A.\ D.\ Martin$^a$, R.\ G.\ Roberts$^b$, M.\ G.\ Ryskin$^{a,c}$
and W.\ J.\ Stirling$^{a,d}$ \\

\vspace*{0.5cm}
$^a \; $ {\it Department of Physics, University of Durham,
Durham, DH1 3LE }\\

$^b \; $ {\it Rutherford Appleton Laboratory, Chilton,
Didcot, Oxon, OX11 0QX}\\

$^c \; $ {\it Laboratory of 
Theoretical Nuclear Physics, St.~Petersburg Nuclear Physics
Institute, Gatchina, St.~Petersburg 188350, Russia}\\

$^d \; $ {\it Department of Mathematical Sciences, University of
Durham, Durham, DH1 3LE }
\end{center}

\vspace*{1.5cm}
\begin{abstract}
We present a formulation which allows heavy quark $(c, b,
\ldots)$ mass effects to be explicitly incorporated in both the
coefficient functions and the splitting functions in the parton
evolution equations.  We obtain a consistent procedure for
evolution through the threshold regions for $c\overline{c}$ and
$b\overline{b}$ production in deep inelastic scattering, which
allows the prediction of the charm and bottom quark densities. 
We use the new formulation to perform a next-to-leading order
global parton analysis of deep inelastic and related hard
scattering data. We find that the optimum fit has
$\alpha_S(M_Z^2) = 0.118$. We give predictions for the charm components of
the proton structure functions $F_2$ and $F_L$ as functions of
$x$ and $Q^2$ and, in particular,
find that $F_2^c$ is in good agreement with the existing measurements.
  We examine the $Q^2$ range of validity of the
photon-gluon fusion model for $c\overline{c}$ electroproduction. 
We emphasize the value of a precision measurement of the  charm
component $F_2^c$ at HERA.
\end{abstract}

\newpage

\noindent {\large \bf 1.  Introduction}

A very wide range of deep inelastic scattering structure function
data can be  successfully described in terms of universal quark
and gluon 
distributions satisfying DGLAP ($Q^2$) evolution equations.
While the formalism for light quarks (i.e. $m_q \ll
\Lambda_{QCD}$)
is on a sound theoretical footing, the treatment of heavy quarks
(i.e. $m_q \gg \Lambda_{QCD}$) is more problematic.
The reason is that in practice one requires a consistent
description
which includes both the kinematical regions $Q^2 \sim m_q^2$
and $Q^2 \gg m_q^2$. 

The problem of how to treat heavy quark contributions
to deep inelastic structure functions has been widely discussed,
see for example \cite{EIL}.  It has been brought into focus
recently by the very precise $F_2^{ep}(x,Q^2)$ data from HERA. 
Both the H1 and ZEUS collaborations have measured
\cite{H1charm,ZEUScharm} the charm quark component $F_2^c$ of the
structure function at small $x$ and have found it to be a large
(approximately 25\%) fraction of the total.  This is in sharp
contrast to what is found at large $x$,
where typically $F_2^c/F_2 \sim O(10^{-2})$
\cite{EMCcharm}.  Since the HERA $F_2$ data \cite{h1f2,zeusf2}
are a potentially
valuable source of information on the gluon distribution, the
value of $\alpha_S$, and the relation between the
non-perturbative (low $Q^2$) and perturbative (high $Q^2$)
domains, it is important that charm component is treated
correctly.

In this paper we present a new, theoretically consistent method
for calculating the heavy quark contributions to the deep
inelastic electroproduction structure functions $F_2$ 
and $F_L$.\footnote{Note that we only consider here the case
of neutral current deep inelastic scattering. The case
of charged current scattering, e.g. $W^-c \to s$, will be
somewhat different but can in principle be treated using
the same techniques.}
Our main focus is on the charm quark contribution, although our
results apply equally well for bottom and top quarks. The most
important feature of our analysis is that it is applicable both
to the threshold region $Q^2 \sim 4 m_c^2$, where phase space
effects are important, and to the asymptotic region $Q^2 \gg 4
m_c^2$, where the charm quark assumes the role of a massless
parton and the DGLAP resummation of leading $(\alpha_S \ln
Q^2)^n$ contributions is necessary.
  
Before describing our formalism and presenting quantitative
predictions, we briefly review existing techniques for treating
the charm quark contribution to $F_2$.  The most simplistic
approach  is to assume that a probe of virtuality $Q^2$ can
resolve a charm quark pair in the proton sea when $Q^2 \gapproxeq
m_c^2$. Since such pairs originate from
fluctuations of the gluon field, $g \to c\overline{c}$, a
perturbative treatment should be valid as long as $m_c^2 \gg
\Lambda_{QCD}^2$.
As $Q^2$ increases, $O(m_c^2/Q^2)$ corrections to the standard 
DGLAP evolution become less important, and the charm quark  can 
be treated as a (fourth) massless quark. These ideas are embodied
in the `massless parton evolution' (MPE) approach
\bea
\label{eq:a1}
c(x,Q^2) &=& 0 \quad \mbox{for}\quad  Q^2 \leq \mu_c^2 \; ,
\nonumber \\
n_f &=& 3 + \theta(Q^2 - \mu_c^2) \quad \mbox{in}\quad P_{qg}, \;
P_{gg}, \; \beta_0, \ldots \; ,
\eea
where $\mu_c = O(m_c)$. The charm contribution to the structure
function is then
\be
F_2^c(x,Q^2) = \textstyle{\frac{8}{9}} x c(x,Q^2)
\label{eq:a2}
\ee
in lowest order.  This is the approach adopted at NLO in the MRS
(and CTEQ) global parton analyses, with $\mu_c$ chosen to achieve
a satisfactory description of the EMC $F_2^c$ data
\cite{EMCcharm}.  For example, in the MRS(A) analysis \cite{mrsa}
it was found that $\mu_c^2 = 2.7$ GeV$^2$ and that this was to a
good approximation equivalent to taking
\be
2 c (x, Q_0^2) \; = \; \delta {\cal S} (x, Q_0^2)
\label{eq:a3}
\ee
with $\delta \approx 0.02$ at the input scale $Q_0^2 = 4$
GeV$^2$.  That is at the input scale, charm $(c + \overline{c})$
was found to have approximately the same shape as the total quark
sea distribution ${\cal S}$, and moreover to form about 2\% of
its magnitude.  The input parameter $\mu_c^2$ (or equivalently
$\delta$) was chosen to give a good
description of the EMC $F_2^c$ data.

Although phenomenologically successful, the MPE model clearly
cannot give a precise description of the charm contribution in
the threshold region.  Two-body kinematics imply that an on-shell
$c\overline{c}$ pair can be created by photon-gluon fusion (PGF)
provided
\be
W^2 = Q^2 {1-x \over x} \geq 4m_c^2
\label{eq:a4}
\ee
where $W$ is the $\gamma^* g \rightarrow c\overline{c}$
centre-of-mass energy.  That is, at small $x$, $c\overline{c}$
production is not forbidden even at small $Q^2 < \mu_c^2$ where
the MPE approach gives zero\footnote{In
ref.~\cite{mrsr} the MPE model was modified by the
introduction of a smooth \lq smearing' function which gave
a gradual onset of the charm distribution from a low input scale
$Q_0^2$, namely $Q_0^2 = 1$ GeV$^2$.}.  In the PGF approach,
which was used, for example, in refs.\ \cite{GHR,PGF,KMS},
$F_2^c$ is calculated using the exact matrix elements and phase
space for the process $\gamma^* g \to c\overline{c}$. In
leading order in $\alpha_S$ we have
\be
F_2^c (x, Q^2) = \int \: dz~C_g (z, Q^2, \mu^2) \: \frac{x}{z} \:
g \left ( \frac{x}{z}, \mu^2 \right ).
\label{eq:a5}
\ee
Note that the scale $\mu^2$ at which the gluon distribution and
the coupling $\alpha_S$ (in the coefficient function $C_g$) are
evaluated is not specified at leading order, but one might guess
that $\mu^2 = O(m_c^2)$ is appropriate.  We discuss a reasonable
choice of $\mu^2$ in more detail in section 3 together with the
effects coming from next-to-leading order (NLO) corrections. 
In contrast to the situation for massless quarks, 
there is no collinear divergence
in the leading-order $\gamma^* g \to c\overline{c}$ calculation:
the integral over the transverse momentum of the produced
$c\overline{c}$ pair is regulated by the quark mass: $\int
dk_T^2 k_T^2/(k_T^2 + m_c^2)^2$. However, this in turn means that
at very high $Q^2$ the leading-order contribution behaves
as $F_2^c \sim \alpha_S (\mu_c^2) \ln (Q^2/m_c^2)$.  Higher-order
corrections also behave as $(\alpha_S \ln (Q^2/m_c^2))^n$, and
fixed-order perturbation theory breaks down.  In fact these large
logarithms are precisely those which are resummed by the DGLAP
evolution equations.  Thus at large $Q^2$ we have to include the
charm quark as a parton in DGLAP evolution.  The exact
next-to-leading order corrections to the PGF structure
function are known \cite{LVN}. 
Indeed very recently \cite{BMSVN} this leading-twist analysis
has been used  to perform   $(\alpha_S \ln (Q^2/m_c^2))^n$
resummation for $Q^2 \gg m_c^2$. However such an approach
is of course not applicable in the threshold region
$Q^2 \gapproxeq   m_c^2$, which we also wish to study because of the HERA
data for $F_2$ in this domain.

Our goal is to include the charm quark in parton evolution in a
consistent way.  First, in section 2, we discuss how to include
the heavy quark mass in the Altarelli-Parisi splitting kernels in
such a way as not to destroy the original parton interpretation,
that is to ensure energy-momentum conservation etc. 
In order to be consistent with the $\msb$ factorization scheme
 the threshold for the onset of the
charm distribution is $Q^2 =  m_c^2$.
Although in fact we will see that charm partons cannot be resolved in
the timescale ($\Delta t \sim 1/\Delta E \sim 2E/Q^2$) of the
fluctuations of the gluon into a $c \bar c$ pair until $Q^2 > 4 m_c^2$.
In section 3 we
discuss the coefficient functions.  The PGF contribution will be
included in the coefficient function for the gluon distribution. 
Thus below the resolution threshold of the charm distribution, $Q^2 <
4 m_c^2$, our result for $c\overline{c}$ production will not be
zero but will agree with the PGF approach.  However, at large
$Q^2$, as was noted in ref.\ \cite{WKT}, part of the PGF cross
section is automatically generated by the evolution of the charm
distribution.  To avoid double counting we must therefore
subtract from the coefficient function given by PGF the
contribution which is generated by evolution in this way.  As a
consequence, above the charm threshold a smaller and smaller
fraction of $F_2^c$ will come from the direct photon-gluon fusion
mechanism, and instead the main part will be generated by
conventional parton evolution.  In section 4 we use the new
formulation to perform a NLO global analysis of deep
inelastic and related hard scattering data.  We find an excellent
overall description with, in particular, a significant charm
component of $F_2$ in the HERA regime.  The analysis allows us to
predict universal charm and bottom quark distributions, $c (x,
Q^2)$ and $b (x, Q^2)$.  In section 5 we present the partonic
decomposition of $F_2^c$ as a function of $Q^2$ and, for
completeness, compare the PGF model estimates.  We also give
predictions for $F_2^b$.  In section 6 we study the charm
component of the longitudinal structure function $F_L$.  Finally,
in section 7, we give our conclusions. \\

\medskip
\noindent {\large \bf 2.  The effects of the charm mass on
evolution}

As mentioned above, our aim is to develop the appropriate
formalism to describe deep inelastic scattering which
incorporates the production of a heavy quark pair (which for
definiteness we take to be $c\overline{c}$) and which allows a
universal charm parton distribution to be obtained from an
analysis of these and other data.  We can identify the charm mass
effects in the structure functions $F_{2,L}^c$ which describe
such scattering from the following subset of
integrations\footnote{The general structure of the integrands is
$k_T^2$/(propagator)$^2$, which for massless partons $\sim
1/k_T^2$.  The $k_T^2$ in the numerator arises from the spin
structure of the parton vertex.}
\be
\ldots \: \int \: \frac{dk_{Ti - 1}^2}{k_{Ti-1}^2} \: \int \:
\frac{dk_{Ti}^2 \: k_{Ti}^2}{(k_{Ti}^2 + m_c^2)^2} \: \int \:
\frac{dk_{Ti + 1}^2}{k_{Ti + 1}^2} \: \ldots
\label{eq:b1}
\ee
where $k_{Ti}$ are the transverse momenta of the $t$ channel
partons.  The mass of the charm quark enters in the $k_{Ti}^2$
integration which results from the $g \rightarrow c\overline{c}$
transition, see Fig.~1.  For the example of the parton chain
shown in Fig.~1 it appears that $m_c^2$ should also have been
retained in the integration over $k_{Ti + 1}^2$.  However, we
show below that this is only needed at next-to-next-to-leading
order (NNLO) in $\alpha_S$.

First we recall the kinematic regime responsible for the
leading-order (LO) result.  LO evolution corresponds to the
resummation of the
leading logarithm terms, $(\alpha_S \ln Q^2)^n$, which arise when
the $n$ emitted partons have strongly ordered transverse momenta
(\ldots $\ll k_{Ti - 1}^2 \ll k_{Ti}^2 \ll k_{Ti + 1}^2$ \ldots).
If two of the partons were to have comparable transverse momenta,
$k_{Tj} \sim k_{Tj + 1}$, then we would lose a $\ln Q^2$ and obtain
instead a NLO contribution of the form $\alpha_S (\alpha_S \ln
Q^2)^{n - 1}$.  We may write the Altarelli-Parisi splitting
functions
\be
P_{ji} \; = \; P_{ji}^{(0)} \: + \: \alpha_S \: P_{ji}^{(1)} \: +
\: \ldots
\label{eq:b2}
\ee
where $P^{(0)}$ is the LO form and $P^{(1)}$ gives the NLO
correction.  $P^{(1)}$ includes virtual corrections to the vertex
and propagators as well as the possibility of producing a second
\lq $s$ channel' parton with comparable transverse momentum. \\

\noindent {\bf 2.1.  LO evolution with charm}

On the scale of the Altarelli-Parisi evolution in $\ln Q^2$, we
see that to a good approximation
\be
\frac{1}{1 + m_c^2/k_{Ti}^2} \; \approx \; \theta (k_{Ti}^2 -
m_c^2),
\label{eq:b3}
\ee
that is the presence of the charm mass simply cuts out the
contribution from the region $k_{Ti}^2 \lapproxeq m_c^2$.
Indeed we will find, at LO accuracy, that
we have massless three flavour
evolution for $Q^2 <  m_c^2$ and massless four flavour
evolution for $Q^2
>  m_c^2$; that is due to strong ordering $(k_{Ti + 1}^2 \gg
k_{Ti}^2)$ we can neglect the charm mass in the $k_{Ti + 1}^2$
integration of Fig.~1.  Therefore at LO the singlet evolution
equations have the symbolic form
\bea
\label{eq:b4}
\dot{g} & = & P_{gg} \otimes g \: + \: \sum_q~P_{gq} \otimes q \:
+
\: P_{gc} \otimes c \nonumber \\
\dot{q} & = & P_{qg} \otimes g \: + \: P_{qq} \otimes q \\
\dot{c} & = & P_{cg} \otimes g \: + \: P_{cc} \otimes c \nonumber
\eea
where $q = u,d,s$ denotes the light quark density functions and
$c$ the charm density.  We have abbreviated $P^{(0)}$ by $P$ and
$\dot{f} = (2 \pi/\alpha_S) \partial f/\partial \ln Q^2$.  At LO
the quark mass effects are simply encapsulated by
\be
P_{ci} \; = \; P_{ci} (m_c = 0) \: \theta (Q^2 - m_c^2)
\label{eq:b5}
\ee
with $i = g$ or $c$, and similarly for $P_{gc}$.  Also the
virtual contribution to $P_{gg}$ must be modified
\be
P_{gg} \; = \; \ldots \: - \frac{1}{3} \: n_f \: \delta (1 - z)
\label{eq:b5a}
\ee
with $n_f = 3 + \theta (Q^2 - m_c^2)$, and, of course, we must
allow for the increase in the number of active flavours $n_f$ in
the running of $\alpha_S$.

Although we show in (\ref{eq:b4}) only the equation for
$\dot{c}$,
we note that each heavy quark $(c, b, \ldots)$ requires a
separate singlet evolution equation \cite{EG} since their
splitting functions have different $\theta$ function
contributions. \\

\noindent {\bf 2.2.  NLO evolution incorporating the charm mass}

At NLO the inclusion of quark mass effects is a bit more
complicated, although it turns out that we only have to take
$m_c$ into account in $P_{cg}$ and then only in the LO part
$P_{cg}^{(0)}$.  (Of course as a consequence we must adjust the
virtual corrections to $P_{gg}$).  The argument is as follows.

We have to improve on approximation (\ref{eq:b3}) of the
$k_{Ti}^2$ integration in (\ref{eq:b1}).  To do this we divide
the integral into two parts
\be
\int \: \frac{dk_{Ti}^2 \: k_{Ti}^2}{(k_{Ti}^2 + m_c^2)^2} \; =
\; \int \: \frac{d (k_{Ti}^2 + m_c^2)}{(k_{Ti}^2 + m_c^2)} \: -
\: \int \: \frac{m_c^2 \: dk_{Ti}^2}{(k_{Ti}^2 + m_c^2)^2}
\label{eq:b6}
\ee
where the first term gives the leading logarithm contribution
that we discussed in section 2.1.  To be specific we have
\be
\int_{k_{Ti - 1}^2}^{Q^2} \: \frac{d (k_{Ti}^2 +
m_c^2)}{(k_{Ti}^2 + m_c^2)} \; = \; \ln \: \frac{Q^2}{m_c^2}
\label{eq:b7}
\ee
for $k_{Ti - 1}^2 \ll m_c^2$, which is equivalent to the
threshold factor $\theta (k_{Ti}^2 - m_c^2)$ of (\ref{eq:b3}). 
The second term in
(\ref{eq:b6}), which is concentrated in the region $k_{Ti}^2 \sim
m_c^2$, gives a constant contribution.  That is, it is a NLO
contribution (containing a factor $\alpha_S$ without an
accompanying $\ln Q^2$).  It means that the $m_c^2$ effects need
only be evaluated in the LO (one-loop) part of the $g \rightarrow
c\overline{c}$ splitting function, $P_{cg}^{(0)}$.  For instance
consider the integration over $k_{Ti + 1}$ of Fig.~1 and the
possibility of $m_c^2$ effects in $P_{cc}$.  Clearly if $k_{Ti +
1}^2 \gg m_c^2$ then the mass terms $m_c^2/k_{Ti + 1}^2$ can be
neglected.  If, on the other hand, $k_{Ti + 1}^2 \sim m_c^2$ then
either $k_{Ti}^2 \ll m_c^2$ and $c (x, k_{Ti}^2) = 0$ or
$k_{Ti}^2 \sim m_c^2$ and we lose two $\ln Q^2$ factors so that
the contribution is NNLO, which we omit here.  That is, at NLO
there are no $m_c$ effects in $P_{cc}$.  A similar argument shows
that this is also true for $P_{gc}$.

In summary, we have shown that at NLO $P_{cc}$ and $P_{gc}$
remain as in section 2.1, whereas
\be
P_{cg} \; = \; P_{cg}^{(0)} (m_c) \: + \: \alpha_S~P_{cg}^{(1)}
(m_c = 0) \: \theta (Q^2 -  m_c^2).
\label{eq:b8}
\ee
That is we need only evaluate the effect of the charm mass on the
LO part of $P_{cg}$.  As a consequence of the change in $P_{cg}$,
we have to adjust the virtual correction to $P_{gg}$ by an amount
\be
\Delta P_{gg}^{(0)} \; = \; - \delta (1 - z) \: \int_0^1 \: dz~z
\biggl (P_{cg}^{(0)} (z, m_c) \: - \: P_{cg}^{(0)} (z, m_c = 0)
\biggr ),
\label{eq:b9}
\ee
see section 2.3.  This adjustment also restores energy-momentum
conservation.

We note that instead of a charm density based on the Renormalization
Group (RG) equations and leading-twist contributions\footnote{The charm density
in the conventional RG approach has been discussed recently in
ref.~\cite{BMSVN}, where the (leading twist) coefficients have been fully
calculated at NNLO in the limit $Q^2 \gg m_c^2$.}
we have introduced an arguably more physical parton density based on the 
leading $\ln Q^2$ decomposition of the Feynman diagrams retaining 
full mass effects.
This charm density is universal\footnote{Of course, to use
the parton density for other processes we must calculate
the coefficient functions
within the same framework.} and the partonic momentum sum rule
is satisfied.
Our definition of the parton density coincides\footnote{In the
$Q^2 \gg m_c^2$ domain the evolution in $Q^2$ is exactly the same
for both definitions, however the initial conditions of such an
evolution, if it were to originate in the threshold region, would be
different; compare our approach with that of ref.~\cite{BMSVN}.}
with the conventional (massless) RG--based definition
for $Q^2 \gg m_c^2$.

We have seen that in our approach at NLO we need only 
consider $m_c \neq 0$ effects in the LO diagrams. It is straightforward
to extend the formalism to allow for charm
mass effects in NNLO evolution.  We need only
consider $m_c \neq 0$ effects in the NLO diagrams. That is
we need to evaluate the
\lq\lq blocks" $gg \rightarrow gg$, $gg \rightarrow
c\overline{c}$, $q\overline{q} \rightarrow c\overline{c}$ to
${\cal O} (\alpha_S^2)$ with $m_c^2$ included explicitly, but
only in the region $k_{Ti}^2 \sim m_c^2$.  For example, for $gg
\rightarrow gg$ we would need to evaluate the diagrams shown in
Fig.~2. \\

\noindent {\bf 2.3.  Evaluation of quark mass effects in
$P_{cg}$}

We note that heavy quark mass effects were studied 
in refs.\ \cite{EG,G} in terms of the anomalous dimensions of the
moments of structure functions.  However, it is difficult to
apply the
results to parton evolution, since in these early studies the
mass correction plays the role of a higher twist contribution. 
As a consequence it violates the sum rules which reflect
energy-momentum and baryon number conservation.

To restore the partonic picture we use \lq\lq old fashioned"
perturbation theory.  That is we calculate the $g \rightarrow
c\overline{c}$ splitting function $P_{cg}$ in the infinite
momentum frame with all three partons on-mass-shell.  The parton
four momenta are shown in Fig.~3.  If the momentum of the gluon
is large, $p_g \gg k_T$ and $m_c$, then the quark momentum is
given
by
\be
k \; = \; \left ( zp_g \: + \: \frac{m_c^2 + k_T^2}{2 zp_g};
\quad
\mbox{\boldmath $k$}_T, zp_g \right ),
\label{eq:b10}
\ee
and similarly for $k^\prime$ with $z \rightarrow 1 - z$ and
$\mbox{\boldmath $k$}_T \rightarrow - \mbox{\boldmath $k$}_T$. 
We may write the probability of the $g \rightarrow c\overline{c}$
splitting in the form
\be
dw_{cg} \; = \; 8g^2 T_R \: \frac{d^2 k_T
dk_{\parallel}}{(2\pi)^3} \; \left [ \frac{1}{(2 zp_g)^2 \: 2 (1
- z)p_g} \right ] \; \frac{{\rm Sp}}{(\Delta E)^2}
\label{eq:b11}
\ee
with colour factor $T_R = \frac{1}{2}$ and where the $[\ldots]$
contain the normalization factors of the two $t$ channel and one
$s$ channel quark lines shown in Fig.~3.  The energy denominators
\be
\Delta E \; = \; E_{c\overline{c}} \: - \: E_g \; = \;
\frac{m_c^2 + k_T^2}{2z (1 - z)p_g}
\label{eq:b12}
\ee
play the role of the quark propagators and the numerator
\bea
\label{eq:b13}
{\rm Sp} & = & \frac{1}{2} \: \delta_{ab}^\perp \: {\rm Tr} \left
( \gamma_a \: \frac{\fmslash{k} + m_c}{2} \; \gamma_b \: \frac{-
\fmslash{k}^\prime + m_c}{2} \right ) \nonumber \\
& & \nonumber \\
& = & (m_c^2 + k_T^2) \: \frac{z^2 + (1 - z)^2}{2z (1 - z)} \; +
\; m_c^2,
\eea
where $\frac{1}{2} \delta_{ab}^\perp$ is the average over the two
transverse polarizations of the (on mass shell) gluon and
$\frac{1}{2} (\fmslash{k} + m_c)$ is the quark density matrix. 
The factor of 8 in (\ref{eq:b11}) arises from the sum over two
polarizations of both the $c$ and $\overline{c}$ and allows for
the $t$ channel parton to be either $c$ or $\overline{c}$.

To identify the splitting function we must rewrite (\ref{eq:b11})
in the form
\be
dw_{cg} \; = \; \frac{\alpha_S}{2\pi} \: \frac{dz}{z} \: \frac{d
Q^2}{Q^2} \: P_{cg}^{(0)} (z, m_c, Q^2)
\label{eq:b14}
\ee
where $dk_{\parallel} = p_g dz$.  The outstanding problem is
therefore to determine the scale $Q^2$ appropriate for $k_T^2$. 
The scale $Q^2$ should be chosen so that it correctly reproduces
the timescale of the fluctuations of the gluon into the
$c\overline{c}$ pair, that is
\be
\Delta t \: \sim \: \frac{1}{\Delta E} \; = \; \frac{2 E_g}{Q^2}
\label{eq:b15}
\ee
where $\Delta E$ is given by (\ref{eq:b12}).  It follows that the
appropriate scale would be
\be
Q^2 \; = \; 2p_g \Delta E \; = \; \frac{m_c^2 + k_T^2}{z (1 - z)} .
\label{eq:b16}
\ee
Indeed this would be the physically natural scale to adopt.
It represents the value of $Q^2$ ($ \gapproxeq 4 m_c^2$) for which the
resolution is sufficient to observe the individual $c$ and $\bar c$
partons within the short $g \leftrightarrow c \bar c$ fluctuation time
$\Delta t$.

Unfortunately (\ref{eq:b16}) is not the conventional scale used in the
$\msb$ factorization scheme in the massless quark limit,
or rather when $Q^2 \gg m_c^2$. To obtain parton densities which
correspond to the $\msb$ scheme for $Q^2 \gg m_c^2$ we must take
the evolution scale to be\footnote{It is connected with the fact that in
dimensional regularization in the $\msb$ scheme a factor
$(k_T^2/\mu^2)^\epsilon$ is introduced into the $dk_T^2/k_T^2$
integration, see for example ref.~\cite{CURCI}. Just as the $\msb$ and
${\rm MS}$ scheme coefficient functions
 differ simply by a constant factor, the coefficient functions in the
$\msb$ scheme and the `natural' scheme based on (\ref{eq:b16}) differ
by a factor which is a function of $z$.}
\be
Q^2 = m_c^2 + k_T^2 .
\label{eq:b16a}
\ee
Of course we could use  (\ref{eq:b16}) as the evolution scale but then
we would have to change the NLO splitting and coefficient functions.
Since the NLO coefficient functions, not only for deep inelastic
scattering but also for other processes, have been calculated  in the
$\msb$ scheme, it is clearly desirable to remain in this scheme.
We therefore adopt (\ref{eq:b16a}) as the evolution scale. Then we can
use the $\msb$ NLO splitting functions $P^{(1)}$ for the massless quarks
and gluons and for the massive quarks for $Q^2 \gg m_c^2$. Moreover for
$Q^2 \gg m_c^2$ all the NLO coefficient functions have the $\msb$ form,
except for one contribution  which we discuss in section~3.2.

Now using the evolution scale (\ref{eq:b16a}) we have
\be
P_{cg}^{(0)} \; = \; 2T_R \: \left [ (z^2 + (1 - z)^2) \: + \:
\frac{2 m_c^2}{Q^2} z(1-z) \right ] \; \theta \left( Q^2  -
m_c^2 \right).
\label{eq:b17}
\ee
Recall that here $P_{cg}$ stands for the splitting into both
$c(\overline{c})$ and $\overline{c}(c)$.  An analogous result for
QED may be found in ref.~\cite{VAK}. The $\theta$ function
represents the threshold $(k_T^2 = 0)$ for generating in the evolution
$c$ and $\bar c$ parton densities, which smoothly tend to the
conventional $\msb$ distributions at high $Q^2$.
We see that even if at small $x$ we have
more than enough energy $W$ to create a $c\overline{c}$ pair,
$W^2 \simeq Q^2/x \gg 4 m_c^2$, then it is possible that
the value of $Q^2$ will be insufficient to resolve the $c$ and $\bar c$
 pair within the short $g \leftrightarrow c \bar c$
fluctuation time $\Delta t$, that is when $Q^2 < 4 m_c^2$.

The complete effect of the quark mass in the NLO splitting
functions which involve the charm quark is contained in
(\ref{eq:b17}).  It leads to the following correction to $P_{gg}$
\be
\Delta P_{gg}^{(0)} \; = \; - \frac{2}{3} \: T_R \: \delta (1 -z)
\;  \frac{ m_c^2}{2 Q^2}  \; \theta (Q^2 - m_c^2),
\label{eq:b18}
\ee
see (\ref{eq:b9}). \\

\medskip
\noindent {\large \bf 3.  Coefficient functions for deep
inelastic charm production}

Just as for light quarks, the contribution of charm to the deep
inelastic structure function $F_2$ is obtained from a convolution
of the parton distributions and the coefficient functions.  We
have
\be
F_2^c (x, Q^2) \; = \; \frac{8}{9} \: \int_x^1 \: dz \:
\frac{x}{z} \: \left [ C_{q = c} (z, Q^2, \mu^2) \: c \left (
\frac{x}{z}, \mu^2 \right ) \: + \: C_g (z, Q^2, \mu^2) \: g
\left (\frac{x}{z}, \mu^2 \right ) \right ]
\label{eq:c1}
\ee
where, due to the quark mass, the coefficient functions have an
explicit dependence on $Q^2$.  The charm quark coefficient
function in (\ref{eq:c1}) has the form
\be
C_c \; = \; C_c^{(0)} \: + \: \frac{\alpha_S}{4 \pi} \: C_c^{(1)}
\: + \: \ldots,
\label{eq:c2}
\ee
while for the gluon we have
\be
C_g \; = \; \frac{\alpha_S}{4\pi} \: C_g^{(1)} \: + \: \ldots \: .
\label{eq:c3}
\ee
At NLO accuracy, to which we are working, we need only the
coefficient functions that are shown explicitly in (\ref{eq:c2})
and (\ref{eq:c3}).

We see that at low scales below partonic threshold, $Q^2 <
m_c^2$, where $c (x, Q^2) = 0$, the structure function $F_2^c$ is
described entirely by $\gamma g$ fusion, that is by the $C_g
\otimes g$ convolution.  However, we will find that as $Q^2$
increases from the charm threshold the contribution from the
$\gamma c$ interaction, $C_c \otimes c$, increases rapidly and
soon becomes dominant.  Of course, as we have already mentioned
in the introduction, when the number of active flavours 
increases from 3 to 4 (as we pass through the threshold region) we must
take care to avoid double counting.  For example, if we were to
take the limit in which charm is regarded as a heavy quark, and
never a parton, then the entire contribution to $F_2$ is
\be
F_2^c \; = \; \frac{\alpha_S}{4\pi} \: C_g^{\rm PGF} \: \otimes
\: g\: .
\label{eq:c4}
\ee
We call this fixed (three) flavour approach the photon-gluon
fusion (PGF) approximation.  From the above discussion it might
appear that the PGF approximation, which clearly gives the
correct NLO answer for $Q^2 <  m_c^2$, will dramatically
undershoot the true prediction as $Q^2$ increases above the charm
resolution threshold.  This is not so, since {\it part} of the Feynman
diagram which is responsible for the important $C_c \otimes c$
parton evolution contribution is contained in $C_g^{\rm PGF}
\otimes g$ in the PGF approximation \cite{WKT}.  Thus to avoid double
counting we will have to subtract this contribution from
$C_g^{\rm PGF} \otimes g$.  The consistent treatment of
charm mass effects will therefore allow us to quantify the
accuracy of the PGF approximation to $F_2^c$ as a function $Q^2$.
\\

\noindent {\bf 3.1.  The charm quark coefficient function for
$F_2^c$}

We must specify the coefficient functions for $F_2^c$ that we
introduced in (\ref{eq:c1})-(\ref{eq:c3}).  First the LO charm
quark coefficient is given by
\be
C_c^{(0)} (z, Q^2) \; = \; z~\delta \biggl (z \: - \: 1/(1 +
m_c^2/Q^2) \biggr ) \; \left (1 \: + \: \frac{4 m_c^2}{Q^2}
\right )
\label{eq:c5}
\ee
where here $z$ is defined with respect to the charm quark
\be
z \; = \; z_0 \; = \; \frac{x}{x^\prime} \; = \; \left (1 \: + \:
\frac{m_c^2}{Q^2} \right )^{-1}.
\label{eq:c6}
\ee
The last equality follows directly from the mass-shell condition
$(x^\prime p + q)^2 = m_c^2$ where $x^\prime$ is the fraction of
the momentum of the proton that is carried by the struck charm
quark, see Fig.~4.  The final factor in (\ref{eq:c5}) allows for
the $F_L$ component of $F_2 = F_T + F_L$ where
\be
\sigma_L/\sigma_T \; = \; 4 m_c^2/Q^2.
\label{eq:c6a}
\ee
Inserting $C_c^{(0)}$ of (\ref{eq:c5}) into (\ref{eq:c1}) gives a
contribution to $F_2^c (x, Q^2)$ proportional to $x c (x^\prime,
Q^2)$ where here the true scale is $\mu^2 = Q^2$.  
In fact at NLO all the $m_c \neq 0$ effects in the charm quark coefficient
function occur in $C_c^{(0)}$. Indeed we justify in section~3.3  that
at  NLO we may simply use the massless quark expression for the
coefficient $C_c^{(1)}$. \\

\noindent {\bf 3.2.  The gluon coefficient function for $F_2^c$}

We may write the gluon coefficient function, defined in
(\ref{eq:c3}), in the form
\be
C_g^{(1)} \; = \; C_g^{\rm PGF} \: - \: \Delta C_g
\label{eq:c7}
\ee
where the PGF expression for $F_2$ is \cite{W}
\bea
\label{eq:c8}
C_g^{\rm PGF} (z, Q^2) & = & \left \{ \left [ z^2 \: + \: (1 -
z)^2 \: + \: \frac{4 m_c^2}{Q^2} \: z (1 - 3z) \: - \:
\frac{8m_c^4}{Q^4} \: z^2 \right ] \right . \: \ln \: \frac{1 +
\beta}{1 - \beta} \nonumber \\
& & \nonumber \\
& & + \; \left . \left [ 8z (1 - z) \: - \: 1 \: - \:
\frac{4m_c^2}{Q^2} \: z (1 - z) \right ] \: \beta \right \} \;
\theta \left ( Q^2 \: \left ( \frac{1}{z} - 1 \right ) \: - \:
4m_c^2 \right ).
\eea
$\beta$ is the velocity of one of the charm quarks in the
photon-gluon centre-of-mass frame
\be
\beta^2 \; = \; 1 \: - \: \frac{4m_c^2 \: z}{Q^2 (1 - z)}.
\label{eq:c9}
\ee
The $\theta$ function in (\ref{eq:c8}), $\theta (W^2 - 4m_c^2)$,
represents the $c\overline{c}$ production threshold, where $W$ is
the c.m.\ energy.  Its presence guarantees $\beta^2 \geq 0$.  The
$\Delta C_g$ term in (\ref{eq:c7}) is necessary to avoid the
double counting of the graph that we have already used to compute
$P_{cg}^{(0)}$, see section 2.3.  That is we must subtract from
$C^{\rm PGF}$ the term $P_{cg}^{(0)} \otimes C_c^{(0)}$ that we
already include in the parton evolution up to $Q^2$.  The $z$
variable in the gluon coefficient functions is defined with
respect to the gluon momentum fraction $x_g$,
\be
z \; = \; \frac{x}{x_g} \; = \; z_0 z^\prime
\label{eq:c10}
\ee
where $z^\prime = x^\prime/x_g$, see Fig.~4.  Thus the explicit
form of the subtraction term is
\bea
\label{eq:c11}
\Delta C_g (z, Q^2) & = & \int \:
\frac{dz_0}{z_0} \: C_c^{(0)} (z_0, Q^2) \: \int_{Q_{\rm
min}^2}^{Q^2} \: d \ln Q^{\prime 2} \: P_{cg} (z^\prime,
Q^{\prime 2}) \nonumber \\
& & \nonumber \\
& = & \left ( 1 \: + \: \frac{4 m_c^2}{Q^2}
\right ) \: \int_{Q_{\rm min}^2}^{Q^2} \: d \ln Q^{\prime 2} \:
P_{cg} (z^\prime, Q^{\prime 2}) .
\eea
The lower limit of integration is given by the \lq\lq 
partonic" $\theta$ function which is hidden in $P_{cg}
(z^\prime, Q^{\prime 2})$, that is
\be
Q_{\rm min}^2 \; = \; m_c^2 .
\label{eq:c12}
\ee
The integration in
(\ref{eq:c11}) may be readily performed to give
\be
\Delta C_g (z, Q^2) \; = \; \left [ \left \{ z^{\prime 2} \: + \:
(1 - z^\prime)^2 \right \} \: \ln \left (
 \frac{Q^2}{m_c^2} \right ) \: + \: z^\prime (1 - z^\prime) \:
\left( 2 -  \: \frac{2 m_c^2}{Q^2} \right)
\right ] \; \left (1 \: + \: \frac{4
m_c^2}{Q^2} \right ),
\label{eq:c13}
\ee
where we require  $Q^2 > Q_{\rm min}^2$ 
and where $z'= z/z_0 = (1+m_c^2/Q^2)z$.

It is interesting to consider the $m_c^2 \rightarrow 0$ limits of
$C_g^{\rm PGF}$ and $\Delta C_g$.  We have
\be
C_g^{\rm PGF} \; \rightarrow \; \left \{ z^2 \: + \: (1 - z)^2
\right \} \; \ln \left( \frac{1-z}{z}\; \frac{Q^2}{m_c^2}
\right) \: + \: 8z (1 - z) - 1
\label{eq:c16}
\ee
as $m_c \rightarrow 0$, which differs from the exact $m_c = 0$
coefficient $C_g^{(1)}$ by the presence of $Q^2/m_c^2$ in the
argument of the logarithm.  However, from (\ref{eq:c13}) we see
that
\be
\Delta C_g \; \rightarrow \; \left \{ z^2 \: + \: (1 - z)^2
\right \} \; \ln \left (  \frac{Q^2}{m_c^2} \right )
\: + \: 2z(1-z)
\label{eq:c17}
\ee
as $m_c \rightarrow 0$, which removes the
$\ln (Q^2/m_c^2)$ term in $C_g^{(1)} = C_g^{\rm PGF} -
\Delta C_g$.

The difference  $C_g^{\rm PGF} - \Delta C_g$ is exactly as in the
conventional (massless) $\msb$ scheme apart from the final term
$2z(1-z)$  in $\Delta C_g$.  This discrepancy is due to the different
order in which the limits are taken in the calculation of the
coefficient function. In the massless case we first take $m_q\to 0$
and then $\epsilon \to 0$, which is equivalent to taking the infrared
cut-off $\Lambda > m_q$, whereas here we have first to take $\epsilon
\to 0$ and then consider the $Q^2 \gg m_c^2$ limit. That is  we
can  only neglect\footnote{If we
were to neglect $m_c^2$  in  (\ref{eq:b17})
 for all $Q^2$ then the additional $2z(1-z)$ contribution in $\Delta C_g$ would
disappear.} $m_c^2$ in $P_{cg}^{(0)}$ of 
(\ref{eq:b17}) if $Q^2 \gg m_c^2$, whereas we need to evaluate
$P_{cg}$ in (\ref{eq:c11}) down to $Q^2 =  Q^2_{\rm min}  = m_c^2$.\\

\noindent {\bf 3.3. Choice of scale}

We now come to the choice of the scale $\mu^2$ in (\ref{eq:c1}).
First we consider the convolution involving the gluon.
The only dependence on the scale $\mu^2$ in the coefficient 
function $C_g$ is in the argument of $\alpha_S$ in 
(\ref{eq:c3}). $C_g^{(1)}$ has no dependence on
$\mu^2$ since all the collinear singularities are regularized by the
heavy quark mass provided that $\mu^2 \lapproxeq m_c^2$.
Nevertheless we have to choose the scale $\mu^2$ for $\alpha_S$
and the gluon distribution. Variation of the scale induces
only NNLO contributions.
There is as yet no 
 complete calculation of the NNLO 
 contributions\footnote{However for $Q^2 \gg m_c^2$
 see \cite{BMSVN}.} (in our framework) 
  applicable for all $Q^2$ which would introduce
terms compensating the variation with scale. We must therefore attempt
to identify the `natural' scale for the process.
We have already
mentioned that the natural scale for the charm convolution is
$\mu^2 = Q^2$.

Due to the different way that the scales enter, the $\alpha_S(\mu^2)
\Delta C_g \otimes g(x/z,\mu^2)$ term does not exactly subtract the LO
charm contribution which comes from the convolution $C_c^{(0)} \otimes
c(x,Q^2)$.  At first loop level the latter term is of the form
\begin{equation}
C_c^{(0)} \int_{m_c^2}^{Q^2} \frac{dq^2}{q^2}\; \alpha_S(q^2)\;
P_{cg}^{(0)}(z) \; g\left(\frac{x}{z},q^2 \right) \; .
\label{eq:Z1}
\end{equation}
Here we also take the natural choice\footnote{Of course we do
not want $\mu^2 < m_c^2$ and so, in the analysis described in section~4,
we take $\mu^2 = {\rm max}\{Q^2,m_c^2\}$.}
 of scale $\mu^2 = Q^2$ in the
$\Delta C_g$ term.  We see that over the whole range of integration in
(\ref{eq:Z1}) we then have $q^2 < \mu^2$ and $\alpha_S(q^2) >
\alpha_S(\mu^2)$. In other words the subtracted value of the $\Delta
C_g$ term is a little less than needed, leading to a lack of
smoothness in $F_2^c$ near threshold. To diminish this effect we could
reduce the scale $\mu^2$ in the $\Delta C_g$ term by taking, say,  
$\mu^2 = \delta Q^2$ with $\delta < 1$. From 
the formal point of view the choice of $\mu^2$ should not matter.
In a NLO analysis it only induces changes at NNLO. However $m_c^2$ is
not so large, and some of the deep inelastic data used in our fit will
be at sufficiently low values of $Q^2$ that we sample scales $\mu^2
\gapproxeq m_c^2$. In this $Q^2$ domain the analysis does have
sensitivity to the choice of scale, showing the need for the NNLO
formulation.

Now let us return to the charm quark coefficient function and, in particular,
explain why the massless approximation is sufficient for $C_c^{(1)}$
at NLO. After the subtraction of the LO contribution (in an analogous
way to (\ref{eq:c7}) for the gluon coefficient function)
the remaining coefficient $C_c^{(1)}$ contains
no $\ln Q^2$ terms. Recall that in the presence of a heavy quark mass we 
lose a logarithm (see (\ref{eq:b6}) and (\ref{eq:b7})). 
Therefore at NLO we need
only consider $m_c \neq 0$ in the LO diagrams. Similarly at NNLO
we only require $m_c \neq 0$ in the NLO graphs and so on.
Therefore the (non-logarithmic ${\cal O} (\alpha_S)$) contribution
$C_c^{(1)}$ is only needed when $Q^2 \gg m_c^2$, as was discussed
in section~2.2 after Eq.~(\ref{eq:b7}).
For $Q^2 \lapproxeq m_c^2$ the charm density $c(x,Q^2) = 0$, while for
$Q^2 \gapproxeq  m_c^2$ two powers of $\ln Q^2$ are lost (one in $P_{cg}$
and one in $C_c^{(1)}$) and so this region contributes only at NNLO.
Thus we can set $m_c = 0$ in $C_c^{(1)}$. Of course we must use the same
definition of the scale $Q^2$, (\ref{eq:b16a}), as for massless evolution
in the $\overline{\rm MS}$ scheme.\\

\noindent {\bf 3.4. The resolution of charm}

Although we now have a definite framework which enables us to incorporate
a $m_c \neq 0$ charm parton into a parton analysis, we immediately
encounter a problem when we confront the data.
The charm density rises rapidly as we evolve up from the threshold
$Q^2 = m_c^2$ (required by the $\msb$ scale (\ref{eq:b16a}))
such that in the region $Q^2 \gapproxeq m_c^2$ it is in conflict
with the data. The reason is clear. The physically reasonable scale at which
the charm parton may be resolved is given by (\ref{eq:b16}) with  a threshold
at $Q^2 = 4 m_c^2$. In order to implement this behaviour we introduce a factor
\begin{equation}
f = \left(1 - \frac{4 m_c^2}{Q^2} \right)\; 
\theta\left(1 - \frac{4 m_c^2}{Q^2} \right)
\label{eq:41a}
\end{equation}
into the charm coefficient function $C_c$ of (\ref{eq:c2}) 
and (\ref{eq:c5}) which,
via (\ref{eq:c11}), then feeds through into $\Delta C_g$. 

At first sight the introduction of the `ad hoc'
factor of $f$ appears to modify even the LO result, which in symbolic
form now may be written
\begin{equation}
F_2^c({\rm LO}) = f C_c^{(0)} \otimes c \; .
\label{eq:cc1}
\end{equation}
This is not so. In fact we will see that the modifications due to the 
introduction of $f$ only enter at NNLO.
At LO we have strong ordering in transverse momenta. LO contributions
therefore only occur for $Q^2 \gg m_c^2$, where $f \to 1$. This reflects the 
fact that mass effects correspond to the loss of
a factor of $\ln Q^2$ and only contribute at the next order of $\alpha_S$.
So the NLO contribution is changed, but only by an amount $\sim \alpha_S
\ln(4 m_c^2 / m_c^2) \sim \mbox{const.}\, \alpha_S$ coming from evolution over
the limited interval $m_c^2 < Q^2 \sim 4 m_c^2$.
However even this contribution is cancelled by the $\Delta C_g$ coefficient function.
To see this we inspect the NLO form
\begin{equation}
F_2^c({\rm LO}+{\rm NLO}) = f \left( C_c^{(0)} + \alpha_S  C_c^{(1)} \right) 
\otimes c \; + \; \alpha_S  C_g^{(1)} \otimes g \; .
\label{eq:cc2}
\end{equation}
To obtain the gluon coefficient function
$C_g^{(1)}$, recall that we evaluated $\gamma g \to  c \bar c$
at ${\cal O} (\alpha_S)$, which we denoted by $\alpha_S C_g^{\rm PGF}$, 
and then subtracted the LO part ($\sim \alpha_S \ln Q^2$), which was
already included, 
\begin{eqnarray}
F_2^c({\rm LO}) & = &f \; C_c^{(0)}
\otimes c  \nonumber \\
& = &f \; C_c^{(0)}\otimes \alpha_S \ln  Q^2 P_{cg}^{(0)} \otimes g 
\equiv \alpha_S \Delta C_g^{(1)} \otimes g \; , 
\label{eq:cc3}
\end{eqnarray}
where it is sufficient to use the LO expression for the charm density.
Thus
\begin{eqnarray}
\alpha_S  C_g^{(1)} & = & \alpha_S \left( C_g^{\rm PGF} -
\Delta C_g^{(1)} \right)   \nonumber \\
&=& \alpha_S C_g^{\rm PGF} - f C_c^{(0)} \otimes\alpha_S \ln Q^2 P_{cg}^{(0)} \; .
\label{eq:cc4}
\end{eqnarray}
The last term of (\ref{eq:cc4}) exactly cancels the first term of 
(\ref{eq:cc2}), so 
\begin{equation}
F_2^c({\rm LO}+{\rm NLO}) = \alpha_S  C_g^{\rm PGF}\otimes g + 
f \alpha_S C_c^{(1)}\otimes c \; .
\label{eq:cc5}
\end{equation}
Thus the introduction of the ad hoc factor $f$ gives rise to a modification
$(1-f)\alpha_S C_c^{(1)} \otimes c$ which only enters at NNLO. Indeed,
in the appendix we show precisely how the  modification is cancelled when 
working to NNLO.

Thus, in summary, in the NLO global parton analysis of section~4
we make the replacement
\begin{equation}
C_c^{(0)} \to f C_c^{(0)} 
\label{eq:41b}
\end{equation}
in (\ref{eq:c5}), and similarly for $C_c^{(1)}$.  Hence
(\ref{eq:c13}) becomes
\begin{equation}
\Delta C_g \to f \Delta C_g \; ,
\label{eq:41c}
\end{equation}
where $f$ is given by (\ref{eq:41a}).
After the introduction of the factor $f$, only the PGF
contribution survives in the region $Q^2 < 4 m_c^2$
below the resolution threshold, even though we have a non-zero
charm parton density for $Q^2 > m_c^2$.
As we evolve above the resolution threshold $Q^2 = 4 m_c^2$
the charm parton component of $F_2^c$ rapidly becomes important.\\
\medskip

\noindent {\large \bf 4.  Charm as a parton in a global analysis}

The measurements of $F_2$ at HERA have become much more precise
with errors as small as $\pm 3\%$ or less.  Moreover, since the
charm component $F_2^c$ of $F_2$ is about 0.25 in the HERA regime
it is important to improve the treatment of charm in the analysis
of deep inelastic scattering data.  This was the objective of
sections 2 and 3 above.  The new formalism incorporates the heavy
quark masses in the parton evolution equations and allows a
determination of the (universal) charm and bottom quark
densities.  Indeed we can predict $c (x, Q^2)$ and $b (x, Q^2)$,
as well as the charm and bottom components of $F_2$, directly from 
a knowledge of the gluon and other quark densities.  There
are no free parameters, although the results do depend on the
values of $m_c$ and $m_b$, and, as usual, on the truncation of
the perturbation expansion.  As in previous analyses, we work to 
NLO.

The new framework is a significant advance on the existing
treatment of charm in deep inelastic scattering.  Recall that two
different types of approach are used at present.  In the first,
charm is set to zero below some scale ($c (x, Q^2) = 0$ for $Q^2
< \mu^2$) and for $Q^2 > \mu^2$ the charm distribution is evolved
assuming that $m_c = 0$.  Although this procedure is clearly
inaccurate in the $c\overline{c}$ threshold region, the parameter
$\mu$ is chosen so that the fixed-target $F_2^c$ data are well
described.  Secondly, we have the PGF approach \cite{PGF,KMS}
based on the calculation of $\gamma^* g \rightarrow
c\overline{c}$ with the correct kinematics, but in which $c$ is
not treated as a parton.  As we have seen, this gives the correct
description of $F_2^c$ for $Q^2 <  m_c^2$ and should remain a
reasonable approximation to $F_2^c$ for $Q^2 \gapproxeq  m_c^2$.
However, the PGF model will inevitably break down at larger $Q^2$
values when charm can no longer be treated as a non-partonic
heavy object and when it begins to evolve more like the lighter
components of the quark sea.

Actually there exists in the literature a third, hybrid, approach
\cite{WKT}. Charm is treated as a new massless parton above
$Q^2 = m_c^2$. That is the $m_c^2$ effects are neglected in the splitting 
functions, although they are included in the coefficient functions to NLO.
This is not quite correct since the neglected $m_c^2$ effects would
give NLO contributions during the evolution.

Before we present our predictions for $c (x, Q^2)$ and $b (x,
Q^2)$, we perform a NLO global analysis of deep inelastic and
related data which incorporates the $m_q \neq 0$ parton evolution
procedure that we introduced in sections 2 and 3.  This may be
regarded as a small refinement of the global analysis
determination of the gluon and light quark densities of ref.\
\cite{mrsr}, but it does allow the gluon (and other parton)
distributions to readjust themselves to accommodate the new
treatment of $c (x, Q^2)$.  Recall that the heavy quark
distributions, $c (x, Q^2)$ and $b (x, Q^2)$, do not contain any
free parameters apart, of course, from $m_c$ and $m_b$. Motivated
by QCD sum rules, we take
$m_c = 1.35$~GeV and $m_b = 4.3$~GeV \cite{PDG96}. We show the 
effects of
varying the value of $m_c$ when we discuss the description of
$F_2^c$.  In fact we find that the
overall description of the data (and in particular of $F_2$ in
the HERA regime) improves compared to our previous
analyses \cite{mrsr}.  The only change to the data set that we
use is the addition of the final NMC data \cite{NMC} for $F_2$. 

We shall present full details of the new global analysis
\footnote{The FORTRAN code for this set of partons, MRRS,
together with the code for computing each flavour component to
$F_1$, $F_2$ and $F_L$ is available by electronic mail from
{\tt W.J.Stirling@durham.ac.uk}, or directly from 
{\tt http://durpdg.dur.ac.uk/HEPDATA/MRS}.} in a future paper in
which we will discuss the improvements of the deep inelastic data
and their implications.  However, in Table~1 we illustrate the
quality of the new fit relative to our previous fit that best
described the HERA data, MRS(R2) \cite{mrsr}. 
  We see that despite now
  having a prescribed charm distribution, the quality of
  the new fit is comparable to or actually slightly better (particularly
  for the small-$x$ $F_2$ measurements) than that of the previous analysis.

\begin{table}[tbh]
\begin{center}
\begin{tabular}{|l|c|rr|} \hline
 Experiment & \# data    &   \multicolumn{2}{|c|}{ $\chi^2$ }  \\
    & &      MRRS & MRS(R$_2)$     \\ \hline
H1~ $F_2^{ep}$   & 193    & 133 &  149  \\
ZEUS~ $F_2^{ep}$     & 204 & 290 & 308   \\  \hline
BCDMS~  $F_2^{\mu p}$    & 174 & 271 & 320   \\
 NMC~  $F_2^{\mu p}$    & 130 & 145 & 134   \\
 NMC~ $F_2^{\mu d}$    & 130 & 119 & 98   \\
E665~ $F_2^{\mu p}$    & ~53  &   60 &  62   \\ 
E665~ $F_2^{\mu d}$    & ~53  &   54 &  60   \\ 
SLAC~ $F_2^{e p}$   & 70 & 96 & 95      \\ \hline
\end{tabular}
\end{center}
\caption{$\chi^2$ values for some of 
the data \protect\cite{h1f2,zeusf2,NMC,bcdms,slac,e665}
used in the global fit.
Note the larger $\chi^2$ values for the E665 points \protect\cite{e665}
than those quoted in ref.\ \protect\cite{mrsr} --- these result from our
correcting our previous incorrect treatment of the E665
experimental errors.}
\label{chisquared}
\end{table}

The HERA data lie in the region where $F_2^c/F_2$ is largest and
there is clear improvement in the new fit for these data. The
value of $\alpha_S$ resulting from the new fit is
$\alpha_S(M_Z^2)$ = 0.118, intermediate to the values 0.113
and 0.120 of MRS(R1) and (R2) and the lower $\chi^2$ for the
BCDMS data in the Table is due to this.  Our prescription for
$\alpha_S(Q^2)$ across charm and bottom thresholds is to 
match the values at $Q^2=m^2_c$, and again at $Q^2 = m^2_b$.
Thus we define
\begin{equation}
\alpha_{S(4)} (Q^2) = \alpha_S (Q^2, 4)
\end{equation}
and take, for 5 flavours,
\begin{equation}
\alpha^{-1}_{S(5)} (Q^2) = \alpha^{-1}_S (Q^2, 5)
+ \alpha^{-1}_S (m^2_b, 4) - \alpha^{-1}_S (m^2_b, 5),
\end{equation}
while for 3 flavours we have
\begin{equation}
\alpha^{-1}_{S(3)} (Q^2) = \alpha^{-1}_S (Q^2, 3)
+ \alpha^{-1}_S (m^2_c, 4) - \alpha^{-1}_S (m^2_c, 3).
\end{equation}

In Fig.~5 we show the flavour decomposition of the sea as a
function of $Q^2$ for two different values of $x$.  Recall that
there are now no input parameters for the heavy quark
distributions, $c (x, Q^2)$ and $b (x, Q^2)$, and that they are
determined in terms of the gluon (and other parton)
distributions.  

We show the description of both the fixed target and HERA data
for $F_2^c$ in the next section.  The charm data are not used in
the global fit.  However, when they become more precise these
data should be included as they will provide a significant extra
constraint on the gluon distribution. 
The gluon density from the new fit compares very closely with
that of MRS(R2). The new gluon is more `valence-like' at $Q_0^2$
= 1 GeV$^2$, but for $Q^2 \geq$ 2 GeV$^2$ both gluon
distributions rise at small $x$ and become increasingly similar
as $Q^2$ continues to increase. \\

\medskip

\noindent {\large \bf 5.  The structure of $F_2^c$}

Fig.~6 shows the partonic decomposition of $F_2^c$ as given by
(\ref{eq:c1}), which may be written in the symbolic form
\be
F_2^c \; = \; C_c~\otimes~c \: + \: C_g~\otimes~g.
\label{eq:e1}
\ee
The gluonic component gives the total production below the charm resolution
threshold, $Q^2 <  4 m_c^2$.  However, the component driven by the
charm distribution rises rapidly above threshold and becomes
dominant at larger $Q^2$.  We also show for comparison the
photon-gluon fusion prediction $C^{\rm PGF} \otimes g$.  The PGF
model and our prediction are identical below threshold, $Q^2 <
4 m_c^2$.  
Above threshold we see that the rapid onset of the charm parton
component $C_c \otimes c$ is largely balanced by the subtraction
$\Delta C_g$  from the PGF result. Let us discuss in turn
the behaviour of $F_2^c$ near the charm threshold and then at large $Q^2$.

The lack of smoothness of $F_2^c$ apparent in Fig.~6 in the charm threshold 
region is due to the mismatch of the subtraction 
term $\alpha_S \Delta C_g \otimes g$
with $C_c^{(0)} \otimes c = C_c^{(0)} \otimes \alpha_S P_{cg}^{(0)} 
\otimes g$ with 
different scales of $\alpha_S$ and $g$ in the two terms, see section~3.3.
In the $\Delta C_g$ term the scale is $\mu^2$ (where we have
taken the natural choice $\mu^2 = Q^2$, or rather 
$ \mu^2 = \mbox{max}\{ Q^2, m_c^2 \}$), 
whereas $\alpha_S$ and $g$ in the second 
term are evaluated at scales varying over the convolution interval $m_c^2$
to $Q^2$. Of course we could have reduced the mismatch by choosing a smaller 
scale $\mu^2$, more representative of the $m_c^2$ to $Q^2$
integration interval. But formally in a NLO
analysis the choice of scale $\mu^2$ (and the mismatch)
should not matter. It  only gives contributions at NNLO.
However in the charm threshold region $\alpha_S(\mu^2)$ is
relatively large and we are sensitive to the choice of $\mu^2$. 
If the NNLO formalism
were available the behaviour  of $F_2^c(x,Q^2)$ in the charm threshold
region would be more stable under variations of $\mu^2$, and 
would have a smoother form in $Q^2$.

As expected these problems evaporate at larger values of $Q^2$. Away from the
charm threshold region ($Q^2 \gapproxeq 20$~GeV$^2$) the predictions for
$F_2^c$ for different $\mu^2$ rapidly approach each other as $Q^2$ increases
and become insensitive to the choice of scale $\mu^2$. The effects of
the {\it evolution} of the charm density are evident. A measure of the effect
is the difference between the prediction of $F_2^c$ (continuous curves
in Fig.~6) and that obtained in the PGF model (dot-dashed curves).
By $Q^2 = 100$~GeV$^2$, for
example, for $x = 0.05$ (0.005) the improved description, in
which charm is treated as a parton, lies some 75\% (30\%) 
above the PGF model.

The comparisons of the predictions for $F_2^c$ with the EMC and
the HERA data are shown in Fig.~7.  The overall agreement over
quite an extensive range of $x$ and $Q^2$ is good.  The dotted
and dashed curves in Fig.~7 show the effect of taking $m_c = 1.2$
and 1.5 GeV respectively, rather than the central value, $m_c =
1.35$ GeV, which we use throughout this paper.

Fig.~8 shows the fraction of charm deep inelastic events as a
function of $Q^2$ for selected values of $x$.  The strong
production of charm at HERA is evident; moreover we see a
sensitive dependence on $x$ and $Q^2$.  If a significant fraction
of the numerous charm events can be cleanly isolated in the
experiments at HERA then the resulting precision measurement of
$F_2^c$, coupled with the measurement of $F_2$, will provide a
powerful double constraint on the gluon distribution, as well as
offering a stringent scheme independent test of QCD along the
lines of that using $F_2$ and $F_L$ proposed by Catani
\cite{CH}.\\

\noindent {\large \bf 6.  Predictions for $F_L^c$}

We may also use the new formalism which incorporates the quark
mass to calculate the charm component $F_L^c$ of the longitudinal
structure function.  We use expressions that are identical to
(\ref{eq:c1})--(\ref{eq:c3}) and (\ref{eq:c7}) but with the
coefficient functions $C_{q = c}$ and $C_g$ that are appropriate
to $F_L^c$.  For the quark coefficient we have
\be
C_c^{(0)} \; = \; \frac{4 m_c^2}{Q^2} \: z~\delta \left (z \: -
\: (1 + m_c^2/Q^2)^{-1} \right ),
\label{eq:f1}
\ee
whereas for $C_c^{(1)}$ we may use the massless quark expression,
since we are working to NLO accuracy.  For the gluon coefficient
for $F_L^c$ we have
\be
C_g^{(1)} \; = \; C_g^{\rm PGF} \: - \: \Delta C_g
\label{eq:f2}
\ee
where
\be
C_g^{\rm PGF} (z, Q^2) \; = \; 4 \beta~z (1 - z) \: - \: 8z^2 \:
\frac{m_c^2}{Q^2} \: \ln~\frac{1 + \beta}{1 - \beta}
\label{eq:f3}
\ee
with $Q^2 > 4 m_c^2 z/(1-z)$, where the quark velocity $\beta$ is
given by (\ref{eq:c9}). Here the subtraction term is
\be
\Delta C_g (z, Q^2) \; = \; \frac{4 m_c^2}{Q^2} \: \biggl [ \; 
\ldots \; \biggr ]
\label{eq:f4}
\ee
where [\ldots] is the expression in the square brackets in
(\ref{eq:c13}).  For $\Delta C_g$ to be non-zero we require $Q^2
> Q_{\rm min}^2$, where $Q_{\rm
min}^2$ is given by (\ref{eq:c12}). Just as for the coefficient functions
for $F_2$, we also include the factor $f$ of (\ref{eq:41a})
in $C_c^{(0,1)}$ and $\Delta C_g$. 

In Fig.~9 we present the predictions for $F_L$ in terms of the
ratio $R^c = F_L^c/F_T^c$.  Due to the factor $4 m_c^2/Q^2$ in
the coefficient function of the LO charm component given in
(\ref{eq:f1}), we have a pronounced peak in $R^c$ just above the
resolution threshold, $Q^2 = 4 m_c^2$. 
 In this region $R^c$ is sensitive
  to the precise choice of the scale $\mu^2$.  As expected $R^c$
  decreases as $Q^2$ increases, as well as becoming more stable
  to changes of scale.  The NLO gluonic component gives a smaller value
of $R^c$ than the charm component.  Hence the peak is more
pronounced at larger $x$ when the gluonic component is less
important.  We also show in Fig.~9 the values of $R = F_L/F_T$.
\\

\noindent {\large \bf 7.  Conclusions}

We have determined the charm and bottom quark densities of the
proton taking into account the effects of their non-zero mass. 
In particular we have presented a formalism which incorporates
$m_c$ and $m_b$ into the Altarelli-Parisi splitting functions and
in the coefficient functions in a consistent way.  We can
therefore evolve up in $Q^2$ taking proper account of the heavy
quark thresholds.  At NLO accuracy we show that the main effect
of the quark mass is in the splitting function $P_{cg}^{(0)}$ (or
$P_{bg}^{(0)}$).

We showed that the threshold for the charm density, $c (x, Q^2)$,
occurs at $Q^2 = m_c^2$.  On the other hand we know that the
threshold for deep inelastic $c\overline{c}$ production is given
by $W^2 = 4 m_c^2$, or equivalently $Q^2 = 4 m_c^2 (1 - x)/x$,
which for small $x$ occurs below the partonic threshold $Q^2 =
m_c^2$.  This apparent contradiction has a simple explanation. 
In the region $Q^2 < 4 m_c^2$ we find that $Q^2$ is too small
to allow sufficient time to observe the $g \rightarrow
c\overline{c}$ fluctuations which occur within the
proton.  Here the photon-gluon fusion mechanism,
$\gamma^* g \rightarrow c\overline{c}$, gives the complete
answer.  For evolution above the partonic 
resolution threshold the structure
of $F_2^c$ is more interesting.  The charm component $\gamma^* c
\rightarrow c$ with a spectator $\overline{c}$ quark (or
vice-versa with $c \leftrightarrow \overline{c}$) increases
rapidly and soon exceeds  the gluonic contribution
$\gamma^* g \rightarrow c\overline{c}$ which only
enters at NLO.  In the partonic
description the LO part of the gluon now has the structure $(g
\rightarrow c\overline{c}) \otimes (\gamma^* c \rightarrow c)$. 
To avoid double counting we must therefore subtract this LO
contribution of the gluon and keep only the part coming from
$C_g^{(1)}$.

In addition to its importance in determining the charm quark
density $c (x, Q^2)$, the correct formulation of charm mass
effects in evolution has become essential in order to obtain an
accurate description of $F_2$ in the HERA domain.  The reasons
are that the charm component of $F_2$ is appreciable ($F_2^c/F_2
\sim 0.25$ for $x \sim 0.001$ and $Q^2 \sim 25$ GeV$^2$) and that
the measurements of $F_2$ at HERA are now much more precise. 

In summary, in this paper we have shown how to treat charm as a parton
for all values of $Q^2$.
The new NLO partonic formulation, which incorporates $m_c \neq 0$
effects, has the following important features.
\begin{enumerate}
\item[(i)] The charm distribution contains no free parameters, except
$m_c$.
\item[(ii)] The partons are universal (that is they can be used in the
NLO description of all hard scattering processes initiated by protons).
\item[(iii)] the splitting and coefficient functions coincide with those
of the (massless) $\msb$ scheme for $Q^2 \gg m_c^2$ (with the one
exception discussed in section~3.2).
\item[(iv)] The momentum and flavour sum rules are conserved.
\item[(v)] There is a definite prescription to enable the formulation
to be extended to include $m_c \neq 0$ effects at NNLO and higher
orders.
\item[(vi)] The new framework, in which the charm density is defined in terms
of a leading $\ln Q^2$ decomposition of the Feynman diagrams retaining
the full mass effects, is applicable in the important 
threshold\footnote{The latter region is
inaccessible to the RG approach to the charm density.
Indeed it is not clear how to formulate
the RG approach when an extra dimensionful parameter ($m_c$) is
essential. For $Q^2 \gg m_c^2$ our formulation reduces to the conventional
RG massless parton approach.} region $Q^2 \gapproxeq  m_c^2$.
\end{enumerate}

Finally, we have used the new prescription to perform
 a global analysis of deep inelastic and related
hard scattering data and generated charm and bottom quark densities.
The analysis {\it predicts} the values of $F_2^c$ 
(and $F_2^b$). We find that the predictions for $F_2^c$ show some sensitivity
to NNLO effects in the charm threshold region ($Q^2 \sim m_c^2$), but become
increasingly stable as $Q^2$ increases above about
 20~GeV$^2$. We find good agreement
with the EMC and H1 measurements of $F_2^c$.
These data, which span a wide range of $(x,Q^2)$, were not
used in the global analysis. Clearly as the experimental precision
increases they should be included, and will impose a valuable
additional constraint in the determination of the parton densities,
and of the gluon in particular.

\medskip
\noindent {\large \bf Acknowledgements}

We thank Valery Khoze, Jan Kwiecinski, Jack Smith  and Robert Thorne
for useful discussions. 
MGR thanks the Royal Society for a Fellowship grant, and for
support from the Russian Fund of Fundamental Research 96 02
17994.

\newpage
\setcounter{equation}{0}
\setcounter{section}{1}
\renewcommand{\theequation}{\Alph{section}\arabic{equation}}
\noindent {\large \bf Appendix}

Here we demonstrate how, if we were to work at NNLO, the factor
$f$ of (\ref{eq:41a}) contributes only at NNNLO. If, for simplicity,
we neglect the light quarks, then we have in analogy to (\ref{eq:cc2})
\begin{equation}
F_2^c(\ldots + {\rm NNLO}) = f\left(
C_c^{(0)} +\alpha_S   C_c^{(1)}  +\alpha_S^2    C_c^{(2)}
\right) \otimes c + \left(
\alpha_S   C_g^{(1)}  +\alpha_S^2    C_g^{(2)} \right)\otimes g \; .
\label{eq:aa1}
\end{equation}
$C_g^{(2)}$ is given by the ${\cal O}(\alpha_S^2)$ expression for the
photon-gluon cross section, $\alpha_S^2 C_g^{(2){\rm PGF}}$ minus
the $\alpha_S^2 \ln^2Q^2$ and $\alpha_S^2\ln Q^2$ contributions which 
are already generated within the LO$+$NLO formalism. That is
\begin{equation}
\alpha_S^2    C_g^{(2)} = 
\alpha_S^2    C_g^{(2){\rm PGF}}  - \Delta C_g^{(2)}
\label{eq:aa2}
\end{equation}
with
\begin{eqnarray}
\Delta C_g^{(2)} & = & f C_c^{(0)} \otimes \left[ 
\left( \alpha_S \ln Q^2 \right)^2 \left( P_{cg}^{(0)} \otimes P_{gg}^{(0)}
+   P_{cc}^{(0)} \otimes P_{cg}^{(0)}   \right)\otimes g
+\alpha_S^2 \ln Q^2 P_{cg}^{(1)} \otimes g
\right]  \nonumber \\
&&+ f \alpha_S C_c^{(1)} \otimes \alpha_S \ln Q^2 P_{cg}^{(0)} \otimes g
+ \alpha_S C_g^{(1)}  \otimes \alpha_S \ln Q^2 P_{gg}^{(0)} \otimes g \; .
\label{eq:aa3}
\end{eqnarray}
Inserting   (A2)  into  (A1) 
and cancelling terms, we find
that the residual ${\cal O}(\alpha_S^2)$ part of $F_2^c$ is
\begin{equation}
F_2^c({\rm NNLO}) = \alpha_S^2    C_g^{(2){\rm PGF}}\otimes g 
 + f  \alpha_S^2   C_c^{(2)}\otimes c \; ,
\label{eq:aa4}
\end{equation}
in analogy to (\ref{eq:cc5}).
The modification $(1-f) \alpha_S^2 C_c^{(2)} \otimes c$ due to the introduction 
of the ad hoc factor $f$ is now of NNNLO.

\newpage
\noindent {\large \bf Figure Captions}
\begin{itemize}
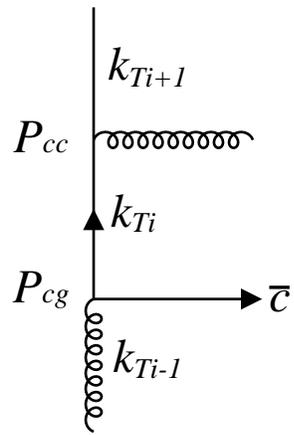
\item[Fig.~1] Part of the parton chain occurring in the
description of deep inelastic scattering which contains the $g
\rightarrow c\overline{c}$ transition.

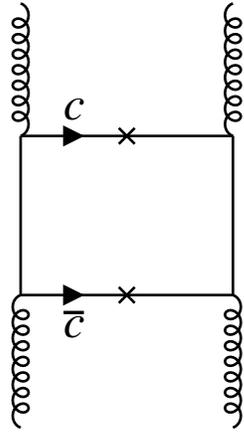
\item[Fig.~2] An example of a \lq\lq block" diagram along the
parton chain, which gives NNLO charm mass effects if the two $s$
channel charm quarks have comparable transverse momenta.  Then
the charm mass should be retained for all the quark lines that
are shown.

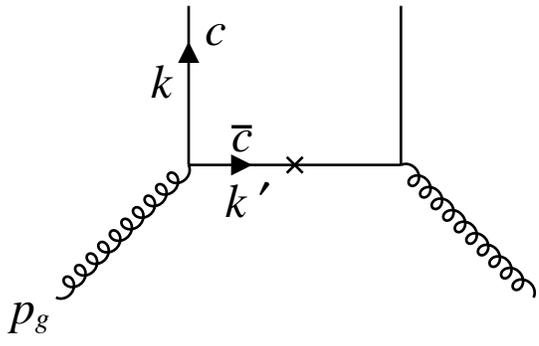
\item[Fig.~3] The diagram used to calculate the charm mass
effects in $P_{cg}^{(0)}$.

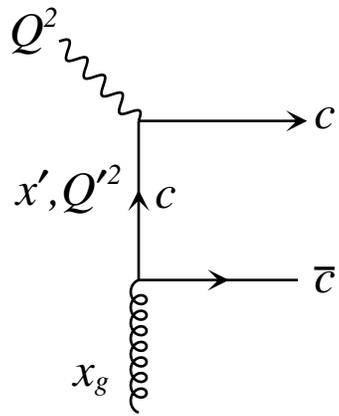
\item[Fig.~4] The variables used in the discussion of the
coefficient functions $C_{q = c} (z, Q^2)$ and $C_g (z, Q^2)$. 
For the charm quark function the variable $z = x/x^\prime$, while
for the gluon function $z = x/x_g$, see eq.\ (\ref{eq:c6}) and
(\ref{eq:c10}) respectively; $x$ is the usual Bjorken $x \equiv
Q^2/2p.q$.

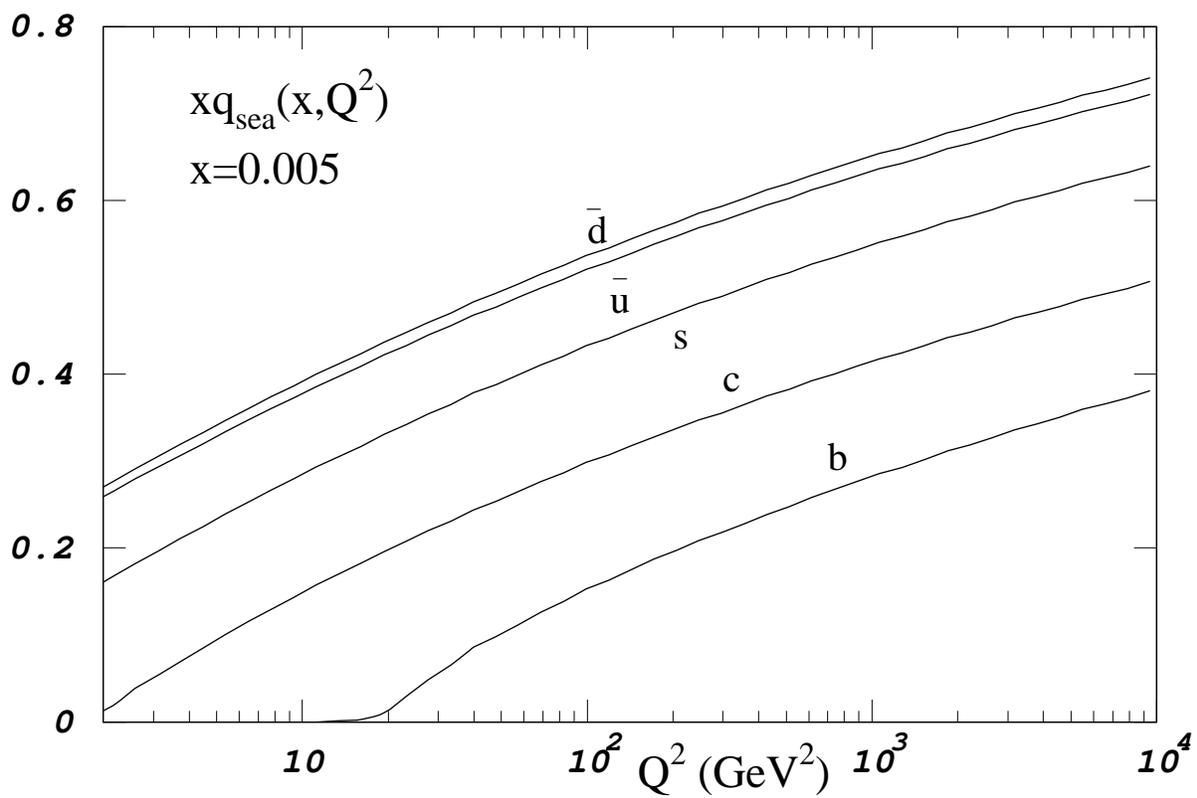
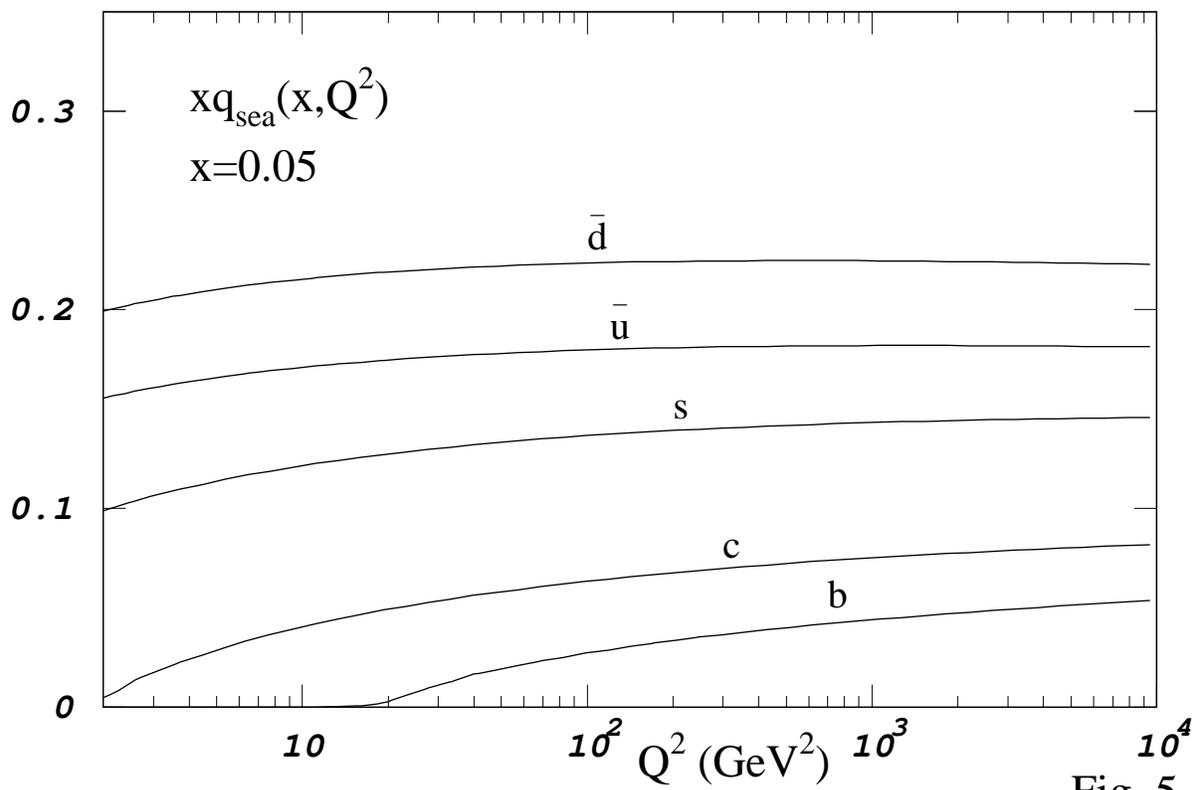
\item[Fig.~5] The flavour decomposition of the quark sea
distribution of the proton as a function of $Q^2$ at two values
of $x$. The total sea
is given by ${\cal S} = 2 (\overline{u} + \overline{d} + s + c +
b)$.

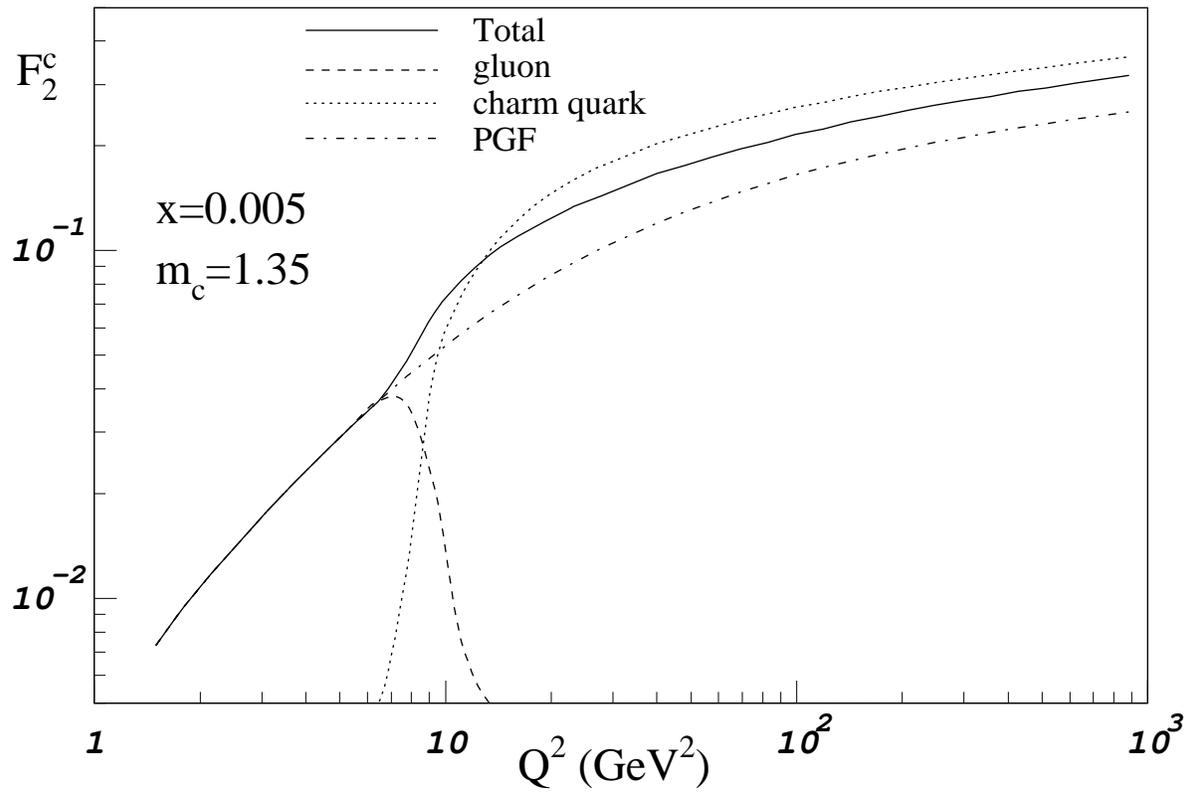
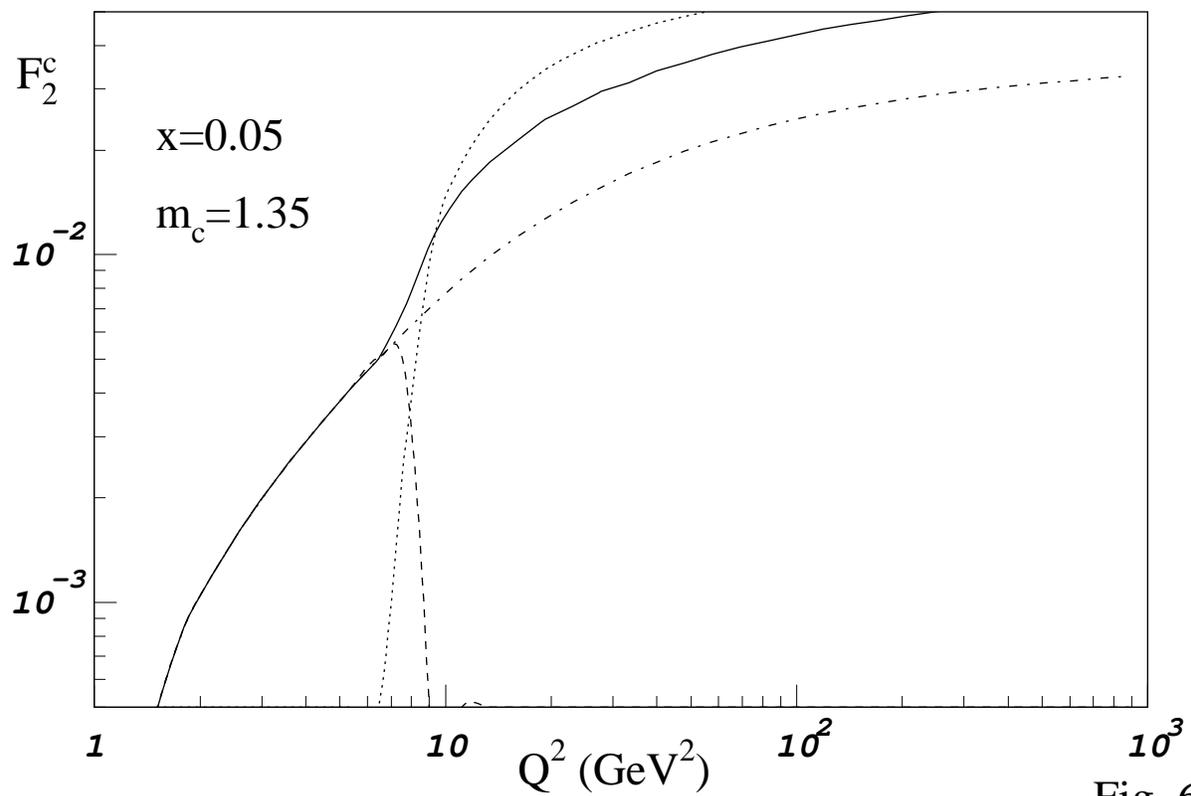
\item[Fig.~6] The partonic decomposition of $F_2^c$ as a function
of $Q^2$ for $x = 0.05$ and $x = 0.005$. For $Q^2 \leq
4m_c^2$ there is only the contribution from $C_g =
C_g^{PGF}$.  For larger $Q^2$, $C_g = C_g - \Delta C_g$ and the
total $F_2^c$ is the sum of this contribution and that from
$C_c$. 

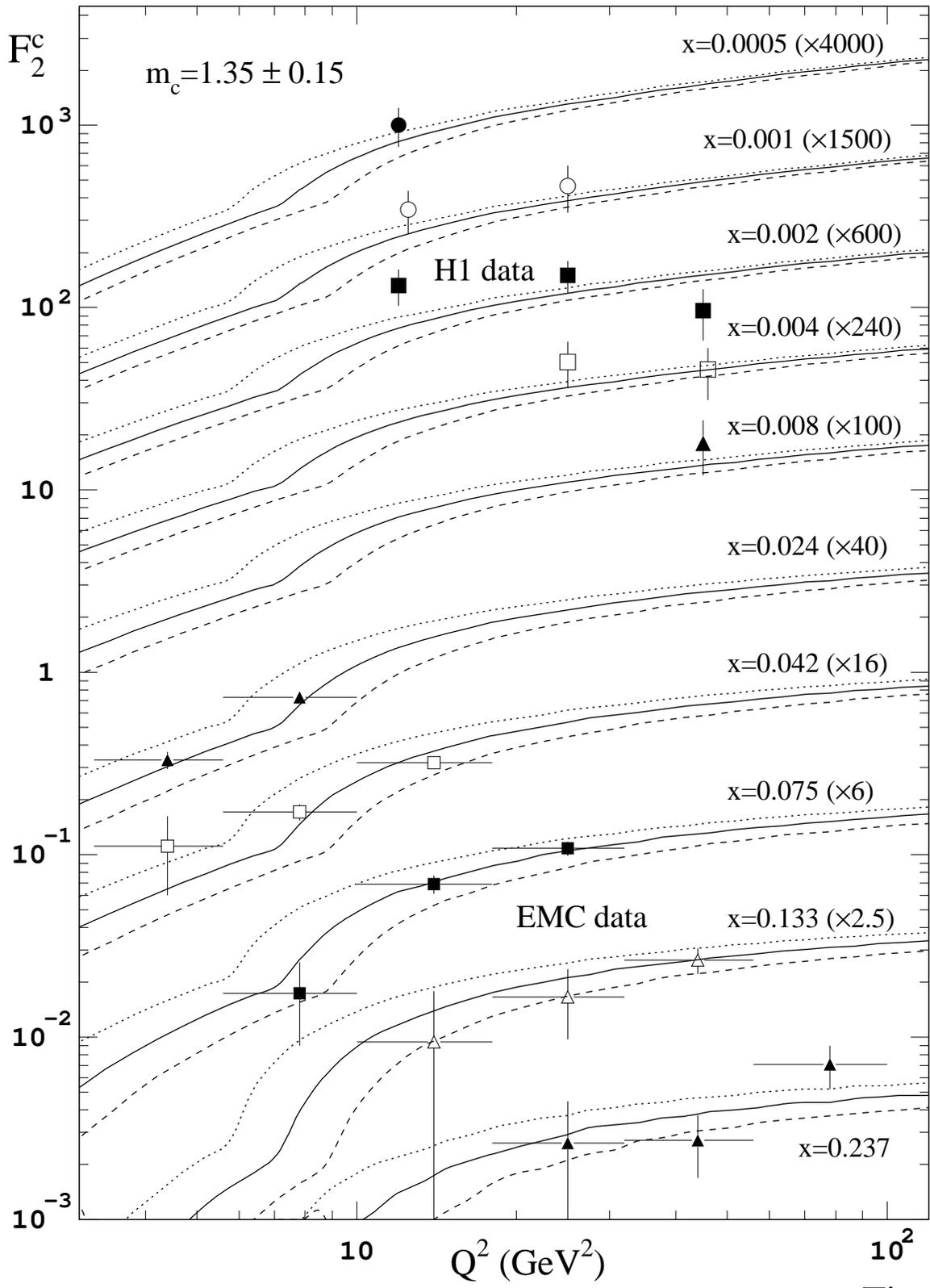
\item[Fig.~7] The description of the EMC and HERA measurements of
$F_2^c$.  The solid line corresponds to our new fit with $m_c$ =
1.35 GeV. The dashed and dotted lines correspond to taking $m_c$
= 1.5 and 1.2 GeV respectively, with all other parameters
unchanged.

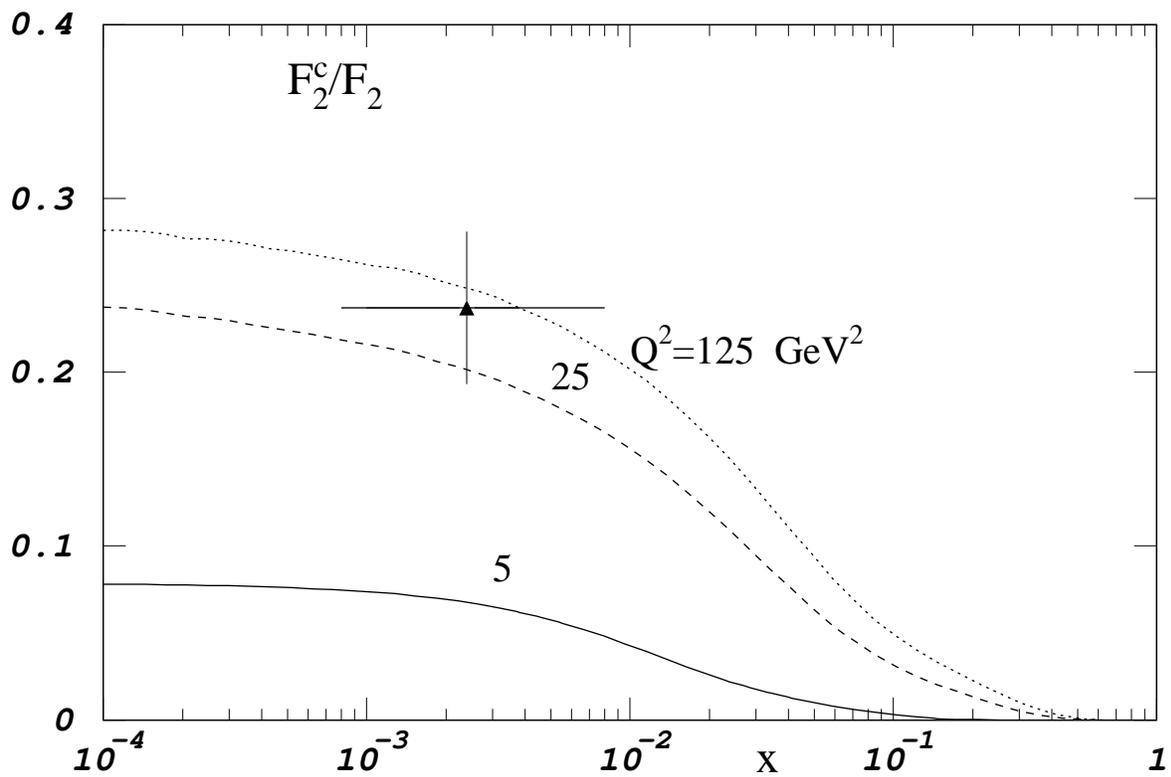
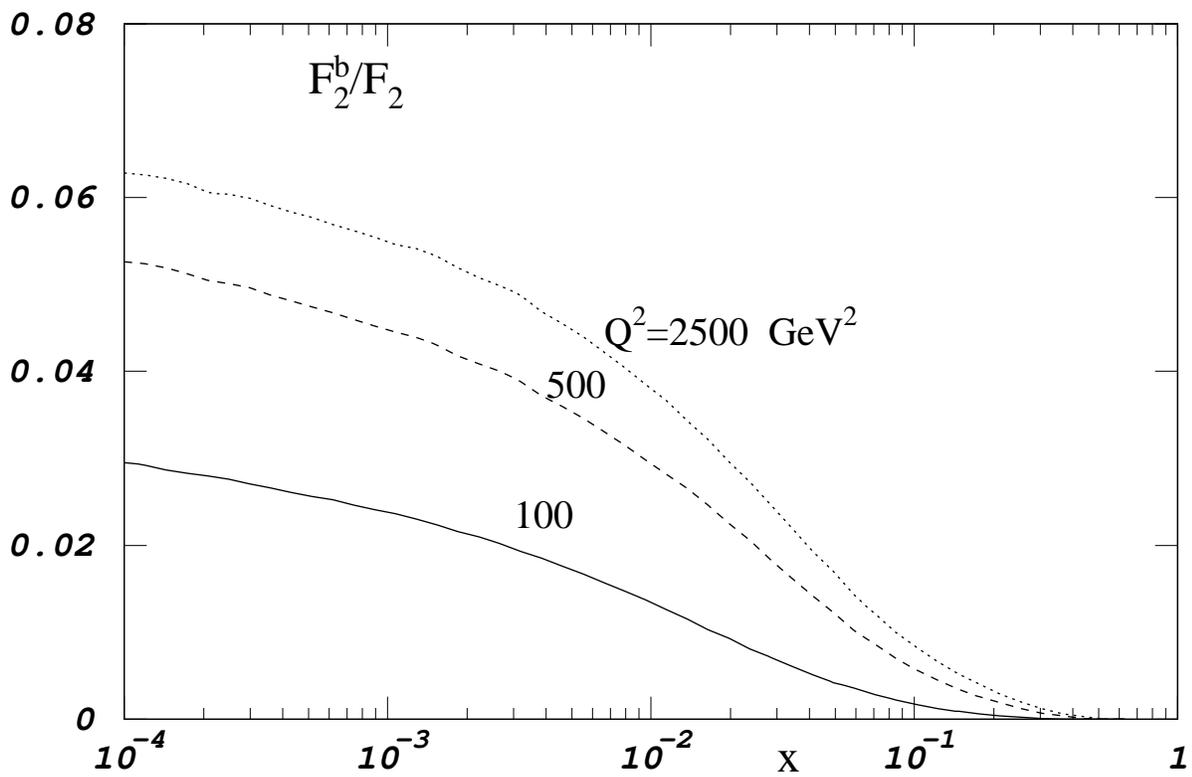
\item[Fig.~8] The ratios $F_2^c/F_2$ and $F_2^b/F_2$ at fixed
values of $Q^2$ resulting from the new global fit (in which we
take $m_c = 1.35$ GeV and $m_b = 4.3$ GeV).  The experimental
data point  shows the estimate from ref.\ \cite{H1charm}
for $F_2^c/F_2$ in the HERA 
kinematic region, $10\; \GeV^2 < Q^2 < 100\; \GeV^2$. 

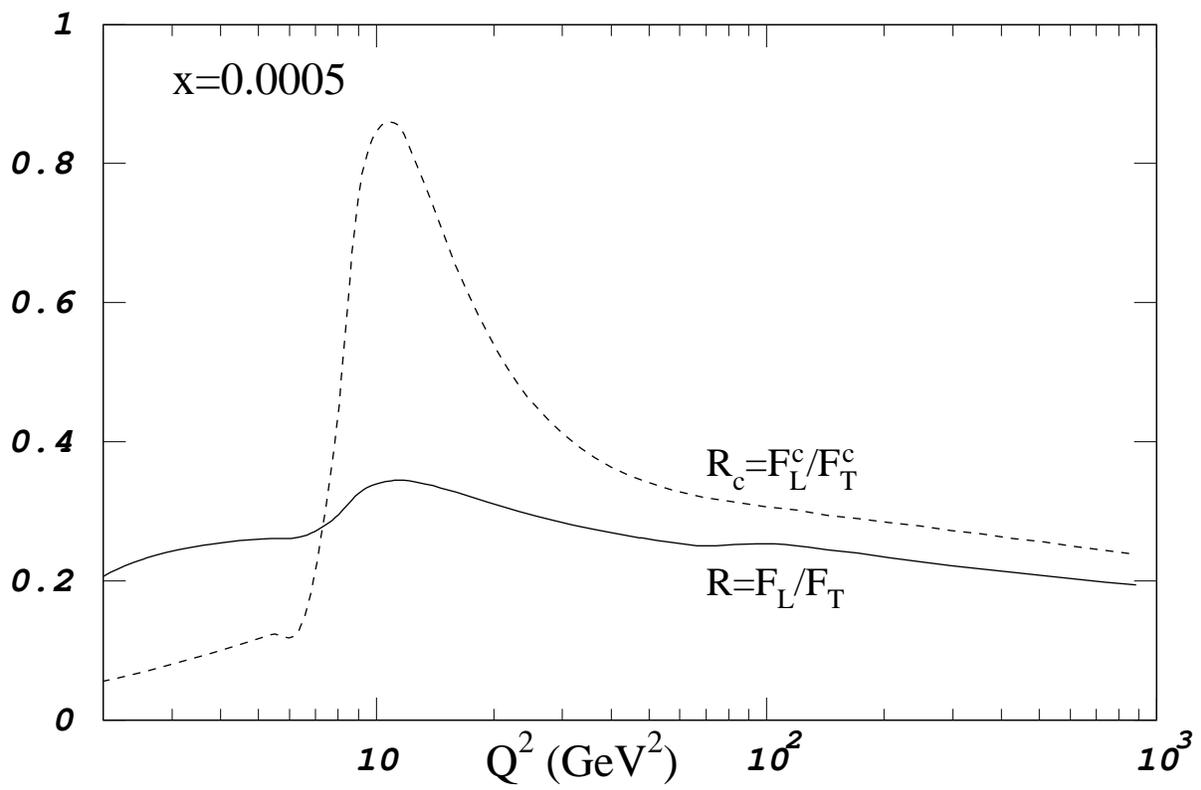
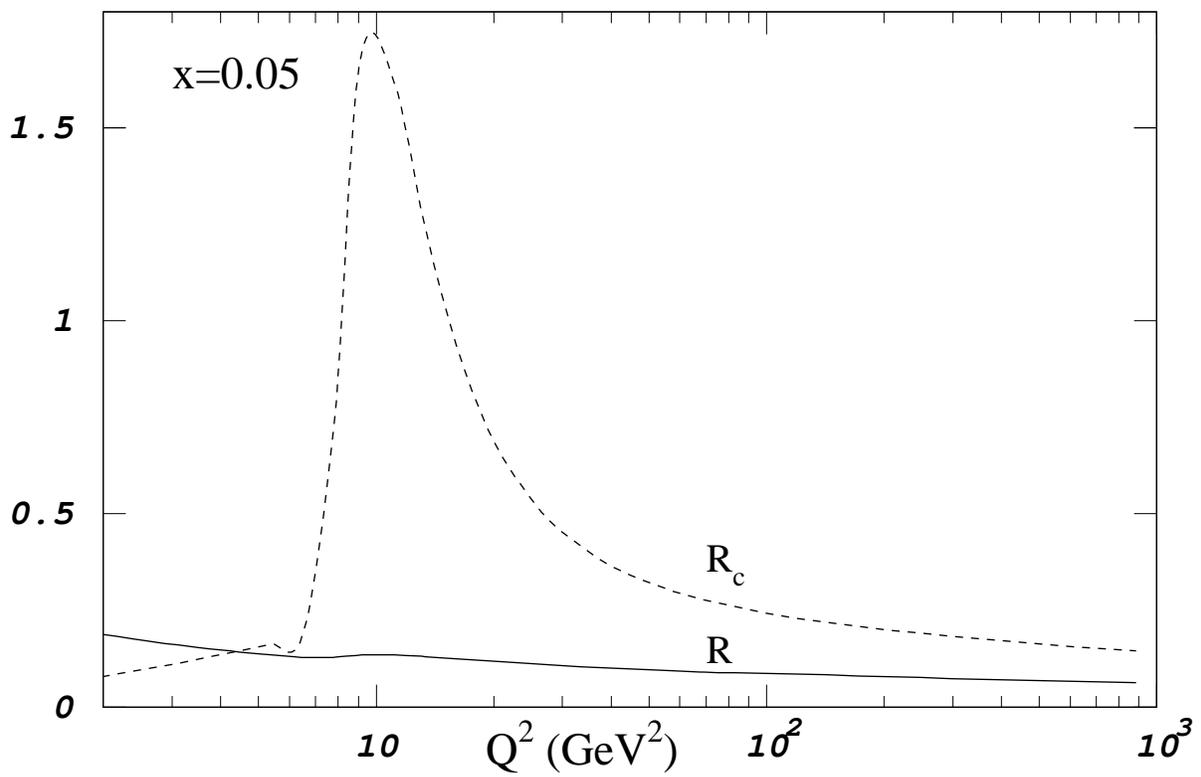
\item[Fig.~9] The predictions for $R^c = F_L^c/F_T^c$ and $R =
F_L/F_T$ as a function of $Q^2$ for $x = 0.0005$ and $x = 0.05$.

\end{itemize} 

\newpage
\begin{thebibliography}{99}

\bibitem{EIL} W.-K.\ Tung, Proc.\ of Int.\ Workshop on Deep
Inelastic Scattering and Related Subjects (DIS 94), Eilat 1994,
Ed.\ A.\ Levy (World Scientific, 1994), p.29.

\bibitem{H1charm} H1 collaboration: C.\ Adloff {\it et al.},
DESY-96-138, July 1996.

\bibitem{ZEUScharm} ZEUS collaboration: ``$D^*$ Production in
Deep Inelastic Scattering at HERA", contribution to the XXVIII
Int. Conf. on HEP, Warsaw 1996.

\bibitem{EMCcharm} EMC collaboration: J.J.~Aubert {\it et al.}, 
Nucl. Phys. {\bf B213} (1983) 31.

\bibitem{h1f2}
H1 collaboration: S.~Aid {\it et al.}, Nucl. Phys.
{\bf B470} (1996) 3.

\bibitem{zeusf2}
ZEUS collaboration: M.~Derrick {\it et al.}, Zeit. Phys. {\bf
C69}
(1996) 607; preprint DESY 96-076 (1996), to be published in Zeit.
Phys.

\bibitem{mrsa} A.D.~Martin, R.G.~Roberts and 
W.J.~Stirling, Phys. Rev. {\bf D50} (1994) 6734.

\bibitem{mrsr} A.D.~Martin, R.G.~Roberts and
W.J.~Stirling, Phys. Lett. {\bf B387} (1996) 419.

\bibitem{GHR} M.\ Gl\"{u}ck, E.\ Hoffmann and E.\ Reya, Z.\
Phys.\ {\bf C13} (1982) 119.

\bibitem{PGF} M.\ Gl\"{u}ck, E.\ Reya and M.\ Stratmann, Nucl.\
Phys.\ {\bf B422} (1994) 37.

\bibitem{KMS} A.J.\ Askew, J.\ Kwiecinski, A.D.\ Martin and P.J.\
Sutton, Phys.\ Rev.\ {\bf D47} (1993) 3775; \\
J.\ Kwiecinski, A.D.\ Martin and P.J.\ Sutton, Z.\ Phys.\ {\bf
C71} (1996) 585.

\bibitem{LVN} E.\ Laernen, S.\ Riemersma, J.\ Smith and W.L.\ van
Neerven, Nucl.\ Phys.\ {\bf B392} (1993) 162.

\bibitem{BMSVN} M.\ Buza, Y.\ Matiounine, J.\ Smith  and W.L.\ van
Neerven, preprint NIKHEF/96-027 (1996).

\bibitem{WKT} M.A.G.\ Aivazis, J.C.\ Collins, F.I.\ Olness and
W.-K.\ Tung, Phys.\ Rev.\ {\bf D50} (1994) 3102.

\bibitem{EG} B.J.\ Edwards and T.\ Gottschalk, Nucl.\ Phys.\
{\bf B196} (1982) 328; \\
Yu.\ P.\ Ivanov, Sov.\ J.\ Nucl.\ Phys.\ {\bf 44} (1986) 317.

\bibitem{G} T.\ Gottschalk, Nucl.\ Phys.\ {\bf B191} (1981) 227.

\bibitem{CURCI} G.\ Curci, W.\ Furmanski and R.\ Petronzio,
Nucl.\ Phys.\ {\bf B175} (1980) 27.

\bibitem{VAK} V.A.\ Khoze {\it et al.}, Nucl.\ Phys.\ {\bf
B65} (1973) 381.

\bibitem{W} E.\ Witten, Nucl.\ Phys.\ {\bf B104} (1976) 445; \\
M.A.\ Shifman, A.I.\ Vainshtein and V.I.\ Zakharov, Nucl.\
Phys.\ {\bf B136} (1978) 157; \\
M.\ Gl\"{u}ck and E.\ Reya, Phys.\ Lett.\ {\bf B83} (1979) 98.

\bibitem{PDG96} {\it Review of Particle Physics}, Particle Data Group,
R.M. Barnett {\it et al.}, Phys. Rev. {\bf D54} (1996) 1.

\bibitem{NMC} NM Collaboration:  M.\ Arneodo {\it et al.}, Nucl.\
Phys.\ (in press), hep-ph-9610231.

\bibitem{bcdms} BCDMS collaboration:  A.C.~Benvenuti {\it et
al.},
Phys. Lett. {\bf B223} (1989) 485.

\bibitem{slac}
L.W.~Whitlow {\it et al.}, Phys. Lett. {\bf B282} (1992)
475.\newline
L.W.~Whitlow, preprint SLAC-357 (1990). 

\bibitem{e665}
E665 collaboration: M.R.~Adams {\it et al.}, Phys. Rev. {\bf D54}
(1996) 3006.

\bibitem{CH} S.\ Catani, to be published in Proc.\ of Workshop on
Deep Inelastic Scattering and Related Phenomena (DIS 96), Rome,
hep-ph/9608310.

\end{thebibliography}
\end{document}